\documentclass[fleqn,usenatbib]{mnras}

\usepackage{newtxtext,newtxmath}

\usepackage[T1]{fontenc}
\usepackage{xcolor}
\usepackage{xcolor}

\newcommand{\LAIA}[1]{} 
\newcommand{\paola}[1]{} 

\DeclareRobustCommand{\VAN}[3]{#2}
\let\VANthebibliography\thebibliography
\def\thebibliography{\DeclareRobustCommand{\VAN}[3]{##3}\VANthebibliography}

\usepackage{graphicx}	
\usepackage{amsmath}	
\usepackage{tabularray}

\title[Open clusters wiggle their tails]{Bar-induced deflection of open cluster tidal tails}

\author[H. Parul et al.]{
Hanna Parul,$^{1}$\thanks{E-mail: anna.parul@obspm.fr}
Laia Casamiquela,$^{1}$
Paola Di Matteo,$^{1}$
Salvatore Ferrone$^{2}$
\\
$^{1}$LIRA, Observatoire de Paris, Université PSL, Sorbonne Université, Université Paris Cité, CY Cergy Paris Université, CNRS, 92190 Meudon, France\\
$^{2}$Université de Strasbourg, CNRS, Observatoire astronomique de Strasbourg, UMR 7550, F-67000 Strasbourg, France\\
}

\date{Accepted XXX. Received YYY; in original form ZZZ}

\pubyear{\the\year{}}

\begin{document}
\label{firstpage}
\pagerange{\pageref{firstpage}--\pageref{lastpage}}
\maketitle

\begin{abstract}
We present a systematic study of how the Galactic bar affects the orientation of tidal tails of open clusters and assess the power of tail morphology to constrain the bar's pattern speed. Using test-particle simulations, we follow the evolution of $\sim 1450$ observed open clusters from the Hunt \& Reffert (2024) catalogue in an axisymmetric reference potential and in eight barred potentials with pattern speeds ranging from $\Omega_b = 20$ km/s/kpc to $\Omega_b = 55$ km/s/kpc. We quantify the bar effect through the deflection angle -- the rotation of the tail orientation in the barred model relative to the axisymmetric case. The deflection angle varies systematically with bar pattern speed and cluster guiding radius.
The largest deflections occur for clusters near the outer Lindblad resonance (OLR), with the sign of the angle set by the orientation of the orbit's pericentre relative to the bar's major axis. For each cluster we measure the distance from the centre beyond which different bar models produce distinguishable tail orientations, and classify each cluster as bar-sensitive or bar-insensitive based on its maximum absolute deflection across the bar models. Comparing with observed tidal tails from the literature, we find that the extended tails of NGC 2632 and the Hyades disfavour moderate pattern speeds. We provide a catalogue of deflection angles, minimal tail extents, and bar-sensitivity flags to guide future observational searches and the re-assessment of existing tidal tail catalogues.
\end{abstract}




\section{Introduction}
The dissolution of open clusters is a complex process influenced by the stellar evolution of their constituent stars, as well as by internal and external dynamical processes all operating over a wide range of time scales \citep[see sections 5 and 6 of][for a recent review]{2024NewAR..9901696C}.  For example, a large fraction of newly formed clusters disperse within the first few Myr due to rapid gas expulsion triggered by supernovae of the cluster’s most massive stars \citep[so-called ``infant mortality'';][]{LadaLada2003, BastianGoodwin2006, BaumgardtKroupa2007, Krumholz2018}. 

For the clusters that survive, we are particularly interested in how their slower dissolution creates stellar streams \citep[see][for a recent review and references therein]{2025NewAR.10001713B}. One driver of such dissolution is two-body relaxation -- an internal process that redistributes energy among member stars, causing some to gain enough energy to exceed the cluster's escape energy and become unbound \citep{gnedin_destruction_1997}.  \footnote{Two-body relaxation only considers the net effect of all stars on a single member, though other dynamical processes occur, such as collisions with binary star systems, which can lead to a single star gaining very high energies. These energies can cause the star to escape from anywhere leading to much more diffuse tidal tails that contribute to a much more diffuse envelope \citep{2024MNRAS.528.5189G}. Some stars can escape with such high velocities that they may even leave the Galaxy \citep{2023ApJ...953...19C}.} Regarding external processes, tidal forces from the Galactic field heat the cluster through tidal shocks, most pronounced during pericentre passages when these forces are strongest \citep{2004AJ....127.2753D}.

In either case, stars escape through the Lagrange points -- in the context of the restricted three-body problem, the locations where the centrifugal force, the Galactic gravitational force, and the cluster's gravitational force balance \citep{2005AJ....129.1906C}. These stars leave with small velocity differences relative to the cluster's centre-of-mass motion. Those that escape from the inner Lagrange point drift ahead, while those that escape from the outer Lagrange point lag behind, leading to the formation of two tidal tails \citep{2004AJ....127.2753D,Just2008, Kupper2008, Kupper2010,2012A&A...546L...7M}. These tidal tails can, to a good approximation, trace the cluster's orbit \citep{2007ApJ...659.1212M, 2024ApJ...967...89I}. Given that stellar streams, to a certain degree, trace the orbit of the host cluster, they have the remarkable utility to infer the properties of the galactic gravitational field \citep{2010ApJ...712..260K,2010ApJ...714..229L, 2011MNRAS.417..198V, 2016ApJ...833...31B}.

While the dynamics of tail formation and their resulting morphology has been extensively studied analytically and through N-body simulations \citep[e.g.][]{Kupper2008, Just2008, Kupper2010, Kupper2012, Chumak2006, Dinnbier20}, observational confirmation of tidal tails around open clusters is recent. In contrast to globular clusters -- whose tails are relatively easy to identify against the sparse stellar halo \citep[e.g.][]{Odenkirchen2001, Grillmair2006, Shipp2018, Malhan2018} -- open clusters evolve within the densely populated Galactic disc and have lower masses, producing tails that are both fainter and embedded in heavy field-star contamination. It was only with the high-precision astrometry from the \textit{Gaia} mission \citep{GaiaDR2} that the discovery of tidal tails of open clusters became possible. 
Concurrently, both \citet{Roser19a} and \citet{Meingast19} reported the first detection of tidal tails associated with an open cluster, namely Hyades, which is the closest open cluster to the Sun.

The detection of the tidal tails of open clusters required dedicated methods. For the Hyades, \citet{Roser19a} used the modified convergent point method (CP), while \citet{Meingast19} analyzed local stars in 3D galactocentric velocity space. \citet{Roser19b} later extended the CP method to NGC~2632 (Praesepe), discovering its tidal tails.
\citet{Tarricq22} applied unsupervised clustering via HDBSCAN in proper motion and parallax space to search for extended structures in 389 local open clusters. Extratidal members of several nearby clusters have been identified with the unsupervised algorithm StarGO, which maps the five-dimensional position–proper-motion space onto a two-dimensional neural network via self-organizing maps \citep{Pang2022, Tang2019, Zhang2020}. \citet{Bhattacharya22} used the ML-MOC algorithm based on k-nearest neighbours (kNN) algorithm and the Gaussian Mixture Models to identify members of the clusters with elongated structures, 20 of which demonstrated the presence of tidal tails. \citet{Vaher23} studied ten young nearby clusters and used traceback computations to assign probabilities for stars to be escapees.

Such methods, however, reliably trace only the inner tails. As \citet{Jerabkova21} showed with N-body simulations, tail stars of older clusters are no longer compact in any phase-space coordinate system and would be missed by clustering-based methods.
To overcome this, \citet{Jerabkova21} introduced the compact convergent point (CCP) method: it uses an N-body model of the cluster to fit the correlation between the velocity difference of the tail stars in the CP method ($v_{\mathrm{||obs}} - v_{\mathrm{||pred}}$) and their spatial distance from the cluster center,  and recover candidate members at much larger distances -- extending the Hyades tails to nearly a kiloparsec. The same method was applied by \citet{Boffin22} to recover long tails of NGC 752. Building on the same idea, \citet{Risbud25} proposed a self-compact convergent point variant that similarly compactifies the velocity space, but without relying on the model obtained from simulations. The largest catalogue of extended tidal tails to date has been produced by \citet{Kos24} who took a different approach: rather than compactifying the velocity space, they constructed the membership likelihood directly from an N-body simulation of cluster dissolution and applied it systematically to 476 open clusters.

Note that the likelihood proposed by \citet{Kos24} depends on the specific Milky Way model chosen, which currently lacks a universal consensus. Many models treat the Milky Way as an axisymmetric, static structure over time and have successfully explained numerous observed phenomena \citep{1991RMxAA..22..255A, 1998MNRAS.294..429D, 2013A&A...549A.137I, galpy, 2017MNRAS.465...76M, Pouliasis17}. When necessary, these models may be extended with time-dependent and non-axisymmetric features, such as spiral arms, a bar, and objects orbiting the Milky Way—including the Magellanic Clouds, molecular clouds, star clusters, and dark matter subhaloes.

Many time-dependent factors may influence the distribution and morphology of open cluster tidal tails. For instance, as these clusters reside in the Galactic disc, \citet{Thomas2023} and \citet{Zhou2026} noted that the Galactic bar can deflect these tails from the cluster’s original orbit and affect their length. They also emphasized that the extent of this effect depends on the bar’s pattern speed, which remains poorly constrained. Different studies, using various tracers and methods, suggest pattern speeds that are slow \citep[$\sim 24$ km/s/kpc,][]{Horta2024}, moderate \citep[$\sim 34-42$ km/s/kpc,][and references therein]{Hunt2025}, or fast \citep[$\sim 54$ km/s/kpc,][]{Ramos2018}. For example, to build their catalogue, \citet{Kos24} included a galactic bar with a pattern speed of 51 km/s/kpc, whereas for the compact convergent method \citet{Jerabkova21} assumed an axisymmetric Milky Way.

Accounting for non-axisymmetric perturbations is therefore important both for reliably identifying tidal tail members in existing and future catalogues, and as an opportunity to provide a novel and complementary constraint on the Galactic bar.

In this work, we present a systematic study of how the Galactic bar affects open cluster tidal tails. \footnote{While this work was being finalised, the independent study of \citet{Kos2026} appeared, presenting a systematic analysis of bar, spiral arms, and GMC effects on open cluster tidal tails using N-body simulations and statistical distance metrics, and reaching broadly consistent conclusions.} We simulate the tidal evolution of $\sim 1450$ observed open clusters from the \citet{HuntReffert24} catalogue in eight barred potentials with pattern speeds spanning $\Omega_b = 20\textup{--}55$ km/s/kpc, as well as in an axisymmetric reference potential, and quantify the deflection of tail orientation relative to the axisymmetric case. We identify which clusters are bar-sensitive -- making them prime targets for constraining the bar's pattern speed -- and which are bar-insensitive and can therefore be reliably modelled under axisymmetric assumptions, setting the basis for a more robust search of tidal tail members in the data. 

The paper is organised as follows: we describe the parameters of the simulations in Sect.~\ref{sec:method} and the sample of open clusters used for analysis in Sect.~\ref{sec:data}. In Sect.~\ref{sec:results} we demonstrate the variety of tail morphologies emerging in the simulations and explore how the deflection of tail orientation varies with cluster orbital parameters and bar pattern speed. In Sect.~\ref{sec:observations} we compare our simulated tails with observed tidal structures from the literature. We discuss the caveats of our study in Sect.~\ref{sec:discussion} and conclude in Sect.~\ref{sec:conclusion}.

\section{Methods}\label{sec:method}

We model the evolution of open clusters using a test particle approach. The simulations are done with python library \texttt{tstrippy}\footnote{\url{https://github.com/salvatore-ferrone/tstrippy}} \citep{Ferrone2023} 
and proceed in two stages. 
First, we compute the orbit of the cluster centre by integrating backward in the Galactic potential from the cluster's present-day coordinates for a time equal to the cluster's age. Second, we integrate a distribution of massless test particles along this computed orbit forward in time from the birth position to the present day. The particles, therefore, evolve in the combined potential of the Galaxy and a time-varying cluster potential that follows the pre-computed orbit. We use a leapfrog integrator with a fixed timestep of $\Delta t = 10^5$ years; therefore, the total number of integration steps varies from $\sim 800$ to $\sim 38000$ depending on the age of the cluster.

The method used in this work has been successfully applied to study the diversity of tidal features in realistic Galactic potentials for a sample of globular clusters \citep{Ferrone2023,Ferrone2025}. The efficiency of the method makes it perfectly suited for systematic study of the evolution of open clusters in barred potentials with different bar pattern speeds.  While the test-particle approach does not produce realistic stellar number counts or density distributions along the tails, it effectively captures how tail morphology -- length and orientation -- responds to variations in the Galactic potential: once a star escapes the cluster, its subsequent trajectory is independent of the potential of the cluster and depends only on mass distribution in the Galaxy.

The following subsections provide the details of the potentials used in the simulations.

\subsection{Galactic potential}\label{sec:gal_pot}

The potential of the Galaxy comprises the axisymmetric part and the rotating bar. 

\begin{equation}
    \Phi_{\mathrm{tot}}(R, z) = \Phi_{\mathrm{axisymm}}(R, z) + \Phi_{\mathrm{bar}}(x, y, z, t)
\end{equation}

The axisymmetric component is adapted from \citet{Pouliasis17} Model II and include spherical dark matter model of the functional form from \citet{AllenSantillan} and two disc components (a thick and a thin disc) represented by Miyamoto-Nagai potentials \citep{MiyamotoNagai1975}:

\begin{equation}
    \Phi_{\mathrm{axisymm}}(R, z) = \Phi_{\mathrm{thin}}(R, z) + \Phi_{\mathrm{thick}}(R, z) + \Phi_{\mathrm{halo}}(r)
\end{equation}

\begin{equation}
    \Phi_{\mathrm{thin}}(R,z) = \frac{-GM_{\mathrm{thin}}}{\left(R^2 + \left[a_{\mathrm{thin}} + \sqrt{z^2 + b_{\mathrm{thin}}}\right]^2\right)^{1/2}}
\end{equation}

\begin{equation}
    \Phi_{\mathrm{thick}}(R,z) = \frac{-GM_{\mathrm{thick}}}{\left(R^2 + \left[a_{\mathrm{thick}} + \sqrt{z^2 + b_{\mathrm{thick}}}\right]^2\right)^{1/2}}
\end{equation}

\begin{equation}
    \Phi_{\rm{halo}}(r) = 
\begin{cases}
\frac{GM_{\rm{halo}}}{a_{\rm{halo}}^{\gamma-1}} \ln\left[\frac{1+\left(r/a_{\rm{halo}}\right)^{\gamma-1}}{1+\left(r_{\rm{cut}}/a_{\rm{halo}}\right)^{\gamma-1}}\right] - \frac{GM_{\rm{total}}}{r_{\rm{cut}}} & r <  r_{\rm{cut}}\\ 
\frac{GM_{\rm{total}}}{r} & r_{\rm{cut}} < r
\end{cases}
\end{equation}
where $r = \sqrt{R^2 + z^2}$. Note that $M_{\rm{halo}}$ is a mass parameter for the halo, and the total mass is given by $M_{\rm{total}} =M_{\rm{halo}} \frac{ \left( r_{\rm{cut}} / a_{\rm{halo}} \right)^{\gamma}  }{1 + \left(r_{\rm{cut}}/a_{\rm{halo}}\right)^{\gamma-1}}$, as given by \citet{1986RMxAA..13..137A}.

For the bar we employ the triaxial Long \& Murali bar \citep{LongMurali92}:
\begin{equation}
\Phi_{\rm bar}(x, y, z) = \frac{GM_{\rm bar}}{2a_{\rm bar}} \ln\left(\frac{x - a_{\rm bar} + T_-}{x + a_{\rm bar} + T_+}\right),
\label{eq:bar_potential}
\end{equation}
where
\begin{equation}
T_{\pm} = \left[(a_{\rm bar} \pm x)^2 + y^2 + \left(b_{\rm bar} + \sqrt{c_{\rm bar}^2 + z^2}\right)^2\right]^{1/2},
\end{equation}
and $a_{\rm bar}$, $b_{\rm bar}$, and $c_{\rm bar}$ are the semi-major, intermediate, and semi-minor axes of the bar.

The values that we used for the potentials are reported in Table~\ref{tab:potential_params}. For the bar pattern speed we explore the set of eight values: 20, 25, 30, 35, 39, 45, 50, and 55 km/s/kpc. This range spans the bulk of literature measurements, which in general converge to the range from 34 to 42 km/s/kpc, with some studies suggesting values as low as 24 or as high as 55 km/s/kpc \citep[][]{Ramos2018, Horta2024, Hunt2018b, Bovy2019, Luccini2024}. Another motivation to include lower pattern speed comes from the spiral arms of the Milky Way, which have similar velocities \citep[e.g.][]{Khalil25}. Although our models include only the bar, the lower end of the $\Omega_b$ range provides general insight into how tails respond to slow non-axisymmetric perturbations.

\begin{table}
\centering
\caption{Parameters of the Galactic potential model. The axisymmetric component is adapted from \citet{Pouliasis17} Model II and includes a spherical dark matter halo \citep{AllenSantillan} and two disc components represented by Miyamoto-Nagai potentials \citep{MiyamotoNagai1975}. The bar is modelled with triaxial Long \& Murali potential \citep{LongMurali92}.}
\label{tab:potential_params}
\begin{tabular}{llc}
\hline
Component & Parameter & Value \\
\hline
\multicolumn{3}{c}{\textit{Axisymmetric Components}} \\
\hline
Dark matter halo & $M_{\rm halo}$ & $2.088 \times 10^{11}$ $M_{\odot}$ \\
 & $a_{\rm halo}$ & 14 kpc \\
                     & $r_{\rm cut}$ & 100 kpc \\
 & $\gamma$ & 2.02 \\                     
\hline
Thin disc & $M_{\rm thin}$ & $3.71 \times 10^{10}$ $M_{\odot}$ \\
 & $a_{\rm thin}$ & 4.8 kpc \\
                 & $b_{\rm thin}$ & 0.25 kpc \\
\hline
Thick disc & $M_{\rm thick}$ & $3.94 \times 10^{10}$ $M_{\odot}$ \\
 & $a_{\rm thick}$ & 2.0 kpc \\
                 & $b_{\rm thick}$ & 0.8 kpc \\
\hline
\multicolumn{3}{c}{\textit{Bar Component}} \\
\hline
Galactic bar & $M_{\rm bar}$ & $2.30 \times 10^{10}$ $M_{\odot}$ \\
 & $a_{\rm bar}$ & 4.0 kpc \\
                      & $b_{\rm bar}$ & 1.0 kpc \\
                      & $c_{\rm bar}$ & 0.5 kpc \\
                      & $\theta_{\rm bar}$ & 28 deg \\
\hline
\end{tabular}
\end{table}

\subsection{Cluster potential}
We use a Plummer model for the cluster potential with initial mass $M_{\mathrm{cluster}} = 5000 M_{\odot}$ and scale radius $a = 3.5$ pc for all clusters: 
\begin{equation}
\label{eq:plummer}
    \Phi_{\mathrm{cluster}} = \frac{-GM_{\mathrm{cluster}}}{\sqrt{r^2 + a^2}}
\end{equation}

The number of particles set to model the stellar distribution is 5000. We use identical initial parameters for all clusters since masses at birth are poorly constrained and since they have minimal impact on tail orientation, which is determined by orbital dynamics of escaped stars in the Galactic potential. The tail length, by contrast, does depend on cluster mass since it sets the velocity dispersion of escaping stars and thereby the rate of tail growth. We therefore focus on tail orientation in our analysis, which is robust to our choice of uniform initial mass.

During the simulation, we consider a particle to have escaped when its kinetic energy relative to the cluster barycenter exceeds the gravitational potential energy due to the cluster (Eq.~\ref{eq:plummer}), and record the timestep at which that occurs as the escape time $t_{\mathrm{esc}}$.

At each timestep, we update the cluster potential by reducing its mass in proportion to the number of escaped particles, thereby incorporating the effect of mass loss on the ongoing tidal disruption. This self-consistent mass prevents formation of cluster remnants that are too compact and does not alter the large-scale tail morphology and tail orientation.

\section{Open clusters: data selection and initial conditions}\label{sec:data}
One of our goals is to provide guidance for future observational searches for tidal tails around open clusters. 
As a starting point, we use the largest and most up-to-date open cluster catalogue from \citet{HuntReffert24}, which lists 5647 open clusters (\texttt{Type == o}).
We use the present-day positions, proper motions, and radial velocities from this catalogue and transform them to Galactocentric reference frame to obtain initial conditions for the backward integration. We integrate for the time equal to the age of the cluster taken from the same catalog. For the transformation between reference frames we assume that the Sun is located at $(x_{\odot}, y_{\odot},z_{\odot}) = (-8.34, 0., 0.027)$ kpc \citep{Chen2001,Reid2014}, the peculiar velocity of the Sun with respect to the local standard of rest (LSR) is $(U_{\odot}, V_{\odot}, W_{\odot}) = (11.1, 12.24, 7.25)$ km/s \citep{Schonrich2010}, and the LSR circular velocity is $v_{\mathrm{LSR}} = 240$ km/s \citep{Reid2014}.
To ensure reliable orbit reconstruction we exclude clusters with radial velocities determined from fewer than five member stars, as unreliable radial velocities lead to large orbital uncertainties that can significantly affect the predicted tail morphology. This filtering yields 2048 clusters. We also removed clusters younger than $\sim 63$ Myr (\texttt{logAge50} $< 7.8$) as they are too young to develop any significant tidal features, which further reduced our sample to 1453 clusters.
Each of those clusters we simulated in axisymmetric potential and eight barred potentials with different pattern speeds, which resulted in $\sim 13000$ simulations.

\section{Results}\label{sec:results} 
Our first result is the dataset of 1453 simulated open clusters evolved in an axisymmetric Galactic potential and eight barred potentials with different bar pattern speeds. The present-day distribution of the simulated particles can be explored through an interactive viewer, which also provides the final-snapshot data for each cluster for download. The viewer, the underlying dataset, and the accompanying catalogue will be made publicly available upon acceptance of this paper. 

In the following subsections, we first provide a general description of simulated tidal tail morphologies (Sect.~\ref{sec:morphology}), then quantify differences in tail orientation between axisymmetric and barred potentials (Sect.~\ref{sec:deflection}), investigate how these deflections vary with cluster location in the Galaxy and bar pattern speed (Sect.~\ref{sec:trends}), and finally relate them to the bar's resonances (Sect.~\ref{sec:resonances}).

\begin{figure*}
    \centering
    \includegraphics[width=2\columnwidth]{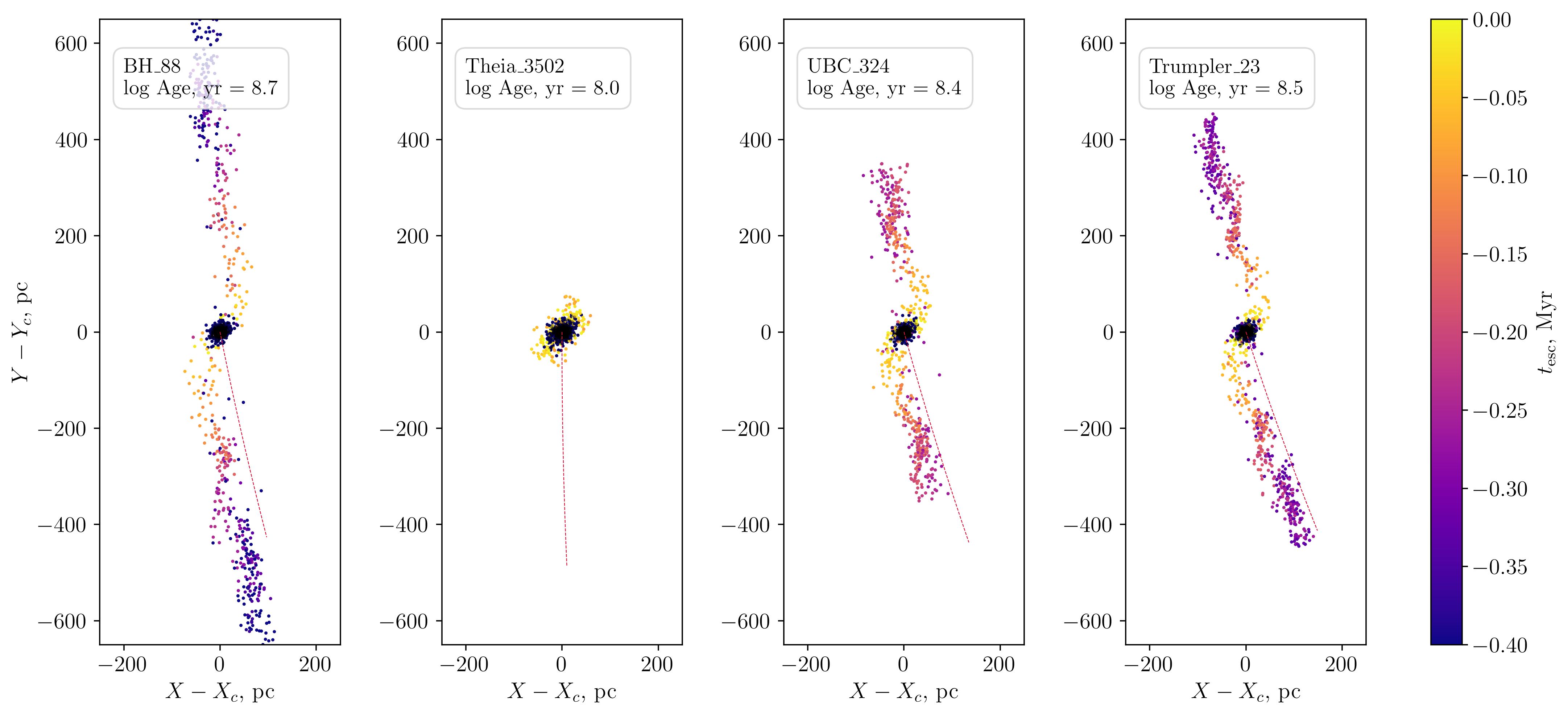}
    \caption{Distribution of test particles clipped to the 99\% enclosed mass fraction for the final snapshot of simulation for four clusters in the axisymmetric potential, shown in the cluster-centered $XY$ plane. Particles are colour-coded by their escape time $t_{\mathrm{esc}}$ and the dashed line shows the local orbital segment. Except for the youngest cluster Theia 3502, all clusters develop tidal tails with an S-shape near the cluster centre, aligning with the orbit at larger distances.}
    \label{fig:shapes_nobar}
\end{figure*}

\begin{figure*}
    \centering
    \includegraphics[width=2\columnwidth]{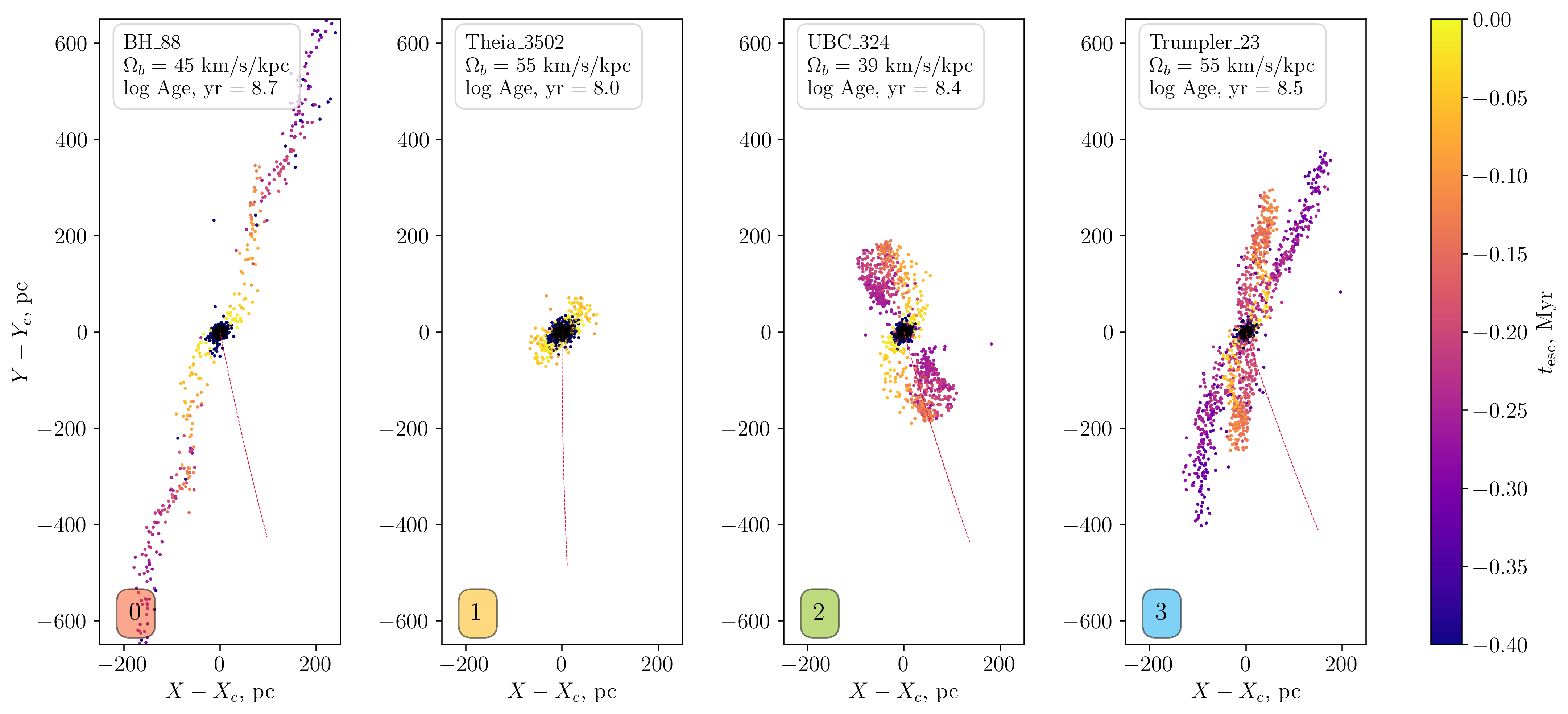}
    \caption{Same four clusters as in Fig.~\ref{fig:shapes_nobar}, each shown in the barred potential at the pattern speed $\Omega_b$ listed in the subpanel. The bar induces a variety of morphological changes relative to the axisymmetric case: BH 88 shows a significant rotation of the tails, UBC 324 develops a shorter, wider and more strongly curved tail, Trumpler 23 demonstrates a multitail structure, while the youngest cluster Theia 3502 remains relatively unchanged. The labels in the lower left corner of each subpanel denote the morphological class assigned by visual inspection. }
    \label{fig:shapes}
\end{figure*}

\subsection{Visual inspection of tail morphology}\label{sec:morphology}
Fig.~\ref{fig:shapes_nobar} showcases four clusters simulated in an axisymmetric potential. They display tidal tails stretched along the orbit of the cluster center, which is showed with dashed line. Near the centre of the cluster the tail has a characteristic S-shape, due to escape of the stars preferentially through Lagrange points. Some tails show clumpy substructure, most clearly visible in Trumpler 23 (last panel), such clumps emerge due to epicyclic movements of the stars in the tails \citep{Just2008,Kupper2008,Kupper2010,Kupper2012}. The length of the tail correlate with the age of the cluster and the width of the tails is similar between different clusters. 

As opposed to regular tails of clusters in an unbarred potential, the tails simulated in a barred Galaxy demonstrate a larger variety of morphologies. Fig.~\ref{fig:shapes} shows the same clusters as in Fig.~\ref{fig:shapes_nobar} evolved in the presence of the bar. While some clusters, like BH 88, retain the regular shape, others, like Trumpler 23, show a multitail distribution, where earlier and later escapees form two distinct structures with different orientation relative to the orbit. 

\begin{figure}
    \centering
    \includegraphics[width=\columnwidth]{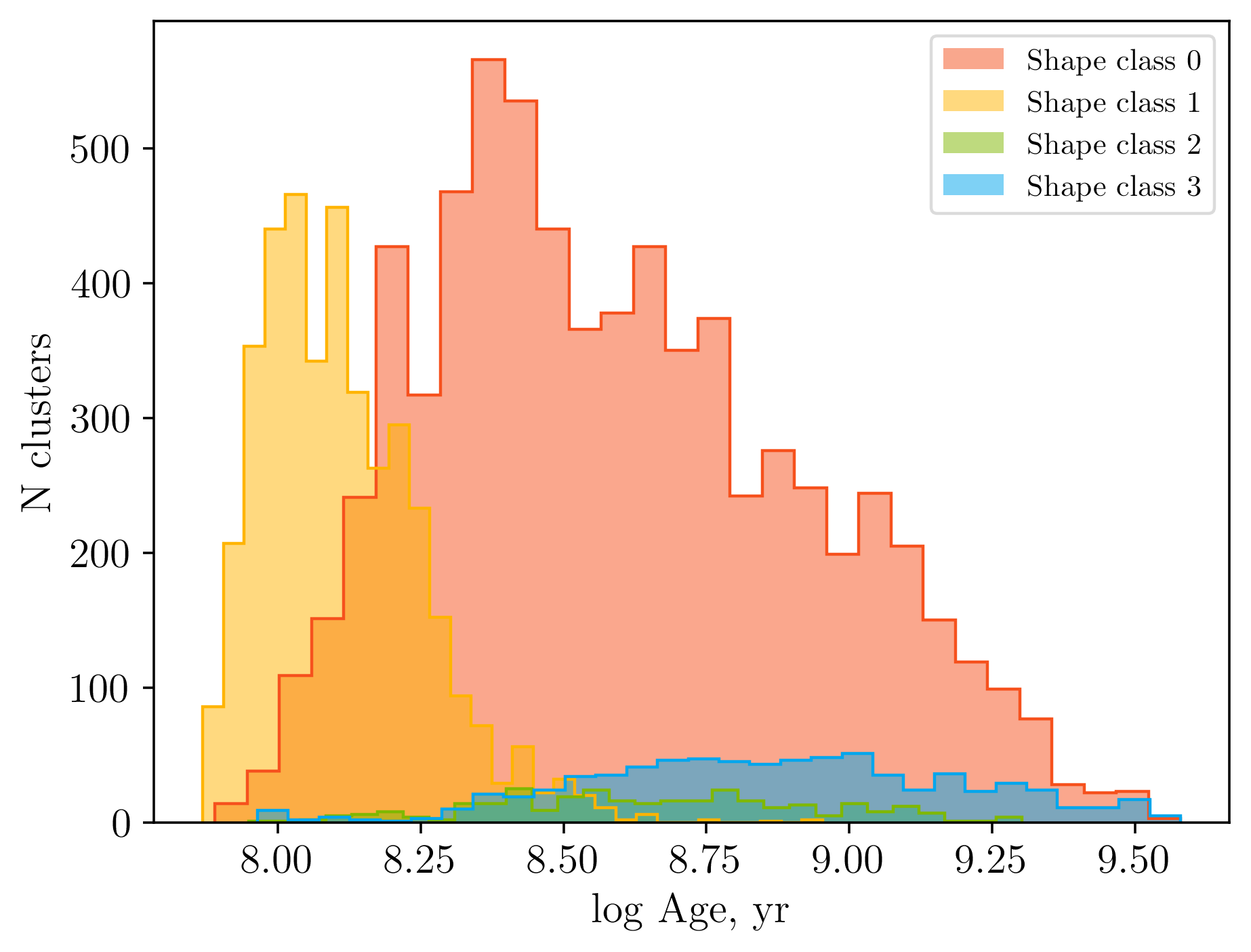}
    \caption{Distribution of cluster ages for each morphological class across all barred simulations. Class \texttt{0} (regular, symmetric tails) dominates at all ages except for the youngest cases. Class \texttt{1} (short, sparse tails) is most prevalent at young ages and largely disappears by \texttt{log Age} $\sim 8.3$ yr. Classes 2 and 3 (irregular and multitail morphologies) are rare and distributed across the wider age range, skewed toward older ages. }
    \label{fig:shapes_age}
\end{figure}

To simplify the subsequent analysis we started with visual inspection of the tail morphologies and identified four main classes based on the appearance of the tails: 
\begin{tblr}{width=\linewidth,colspec={Q[1,l] Q[4,l]}}
Class \texttt{0}  & Constitutes the clusters with regular tails with well-defined orientation, which could be approximated by a curve. For example: BH 88 integrated with $\Omega_b = 45$ km/s/kpc.   \\ 
Class \texttt{1} & Contains the clusters which display very short and dispersed tails, such tails often have a high ratio of width to length. For example: Theia 3502 integrated with $\Omega_b = 55$ km/s/kpc. \\
Class \texttt{2} & Clusters that usually have some sort of preferred orientation, but the width of the stream might be too large, or the ends of the S-shape are too strongly curved. For example: UBC 324 in the Galaxy with $\Omega_b = 39$ km/s/kpc\\ 
Class \texttt{3} & All peculiar cases with multitail distributions and those that lack preferred orientation. For example: Trumpler 23 in the barred potential with $\Omega_b = 55$ km/s/kpc  \\ 
\end{tblr}
Fig.~\ref{fig:shapes} shows the examples for each of the classes with the class label denoted in the lower left corner of each subpanel.  Fig.~\ref{fig:shapes_age} shows the distribution of ages for each of the shapes. As expected, class \texttt{1} is the youngest, peculiar-shaped clusters of classes \texttt{2} and \texttt{3} show a large range of ages, skewed to older values than those of class \texttt{0}. We provide the class label for each cluster in each bar model in the final published catalog.

\subsection{Quantifying tidal tails orientation} \label{sec:deflection}

From comparison of clusters on Fig.~\ref{fig:shapes_nobar} and Fig.~\ref{fig:shapes} it is easy to notice that even clusters retaining regular morphologies (class \texttt{0}) can show significantly different tail lengths and orientations between barred and axisymmetric models. In this work, we do not analyse variations in tail length, since tail extent would be affected by our choice of uniform initial cluster masses. Instead, we focus on tail orientation -- specifically, how tidal tails rotate in the $XY$ plane when evolved in barred versus axisymmetric potentials.

\begin{figure}
    \centering
    \includegraphics[width=\columnwidth]{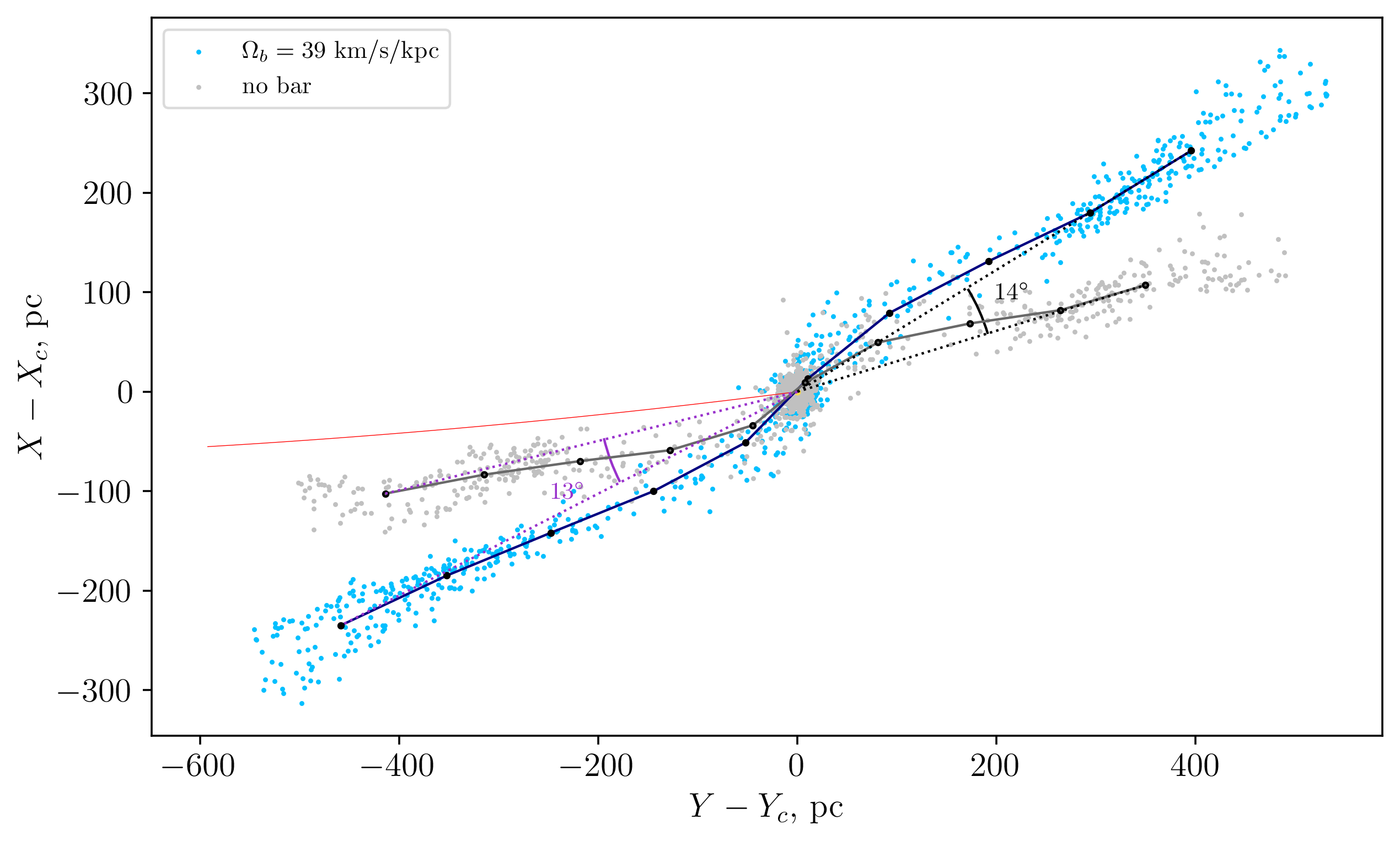}
    \caption{Illustration of the deflection angle measurement for a cluster modelled with $\Omega_b = 39$ km/s/kpc. Grey and blue points show the particle distributions in the axisymmetric and barred models, respectively. The grey and blue curves show the principal curves fitted by \texttt{elpigraph} to the axisymmetric and barred distributions clipped to 99\% enclosed mass. The dotted vectors connecting the cluster centre to the outermost node of each tail are shown for both models, and the deflection angles are measured between them. The dashed red line shows the local orbital segment.}
    \label{fig:angle_measurement}
\end{figure}

To quantify this effect, we measure the deflection angle between tail orientations in the two models. We first identify the spatial extent of each tail by clipping the particle distribution to the 99\% mass fraction, then fitting principal curves to this distribution using \texttt{elpigraph} \citep{elpigraph}, which constructs a graph of nodes tracing the centerline of the stellar distribution (shown in Fig.~\ref{fig:angle_measurement} as grey and blue curves for axisymmetric and barred models, respectively).  

Then we compute the angle between vectors connecting the cluster centre to the outermost node of each tail, measured separately for the leading and trailing tails and averaged. The sign of the angle indicates whether the tail in the barred model is rotated clockwise (positive) or counterclockwise (negative) relative to the axisymmetric case.

\begin{figure}
    \centering
    \includegraphics[width=\columnwidth]{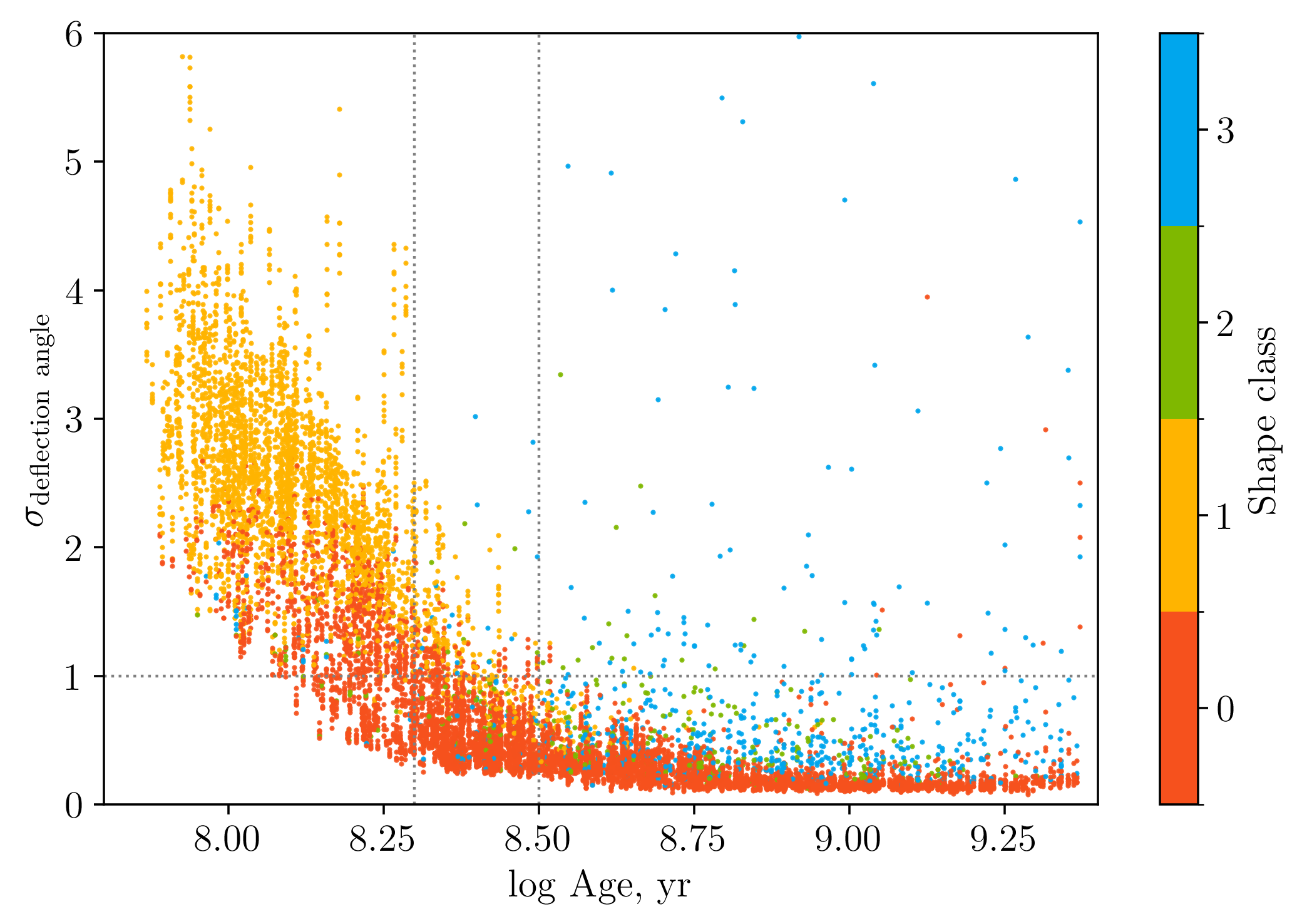}
    \caption{Standard deviation of the bootstrap deflection angle distribution as a function of cluster age, for all barred simulations. Points are colour-coded by morphological shape class. The horizontal dotted line marks $\sigma = 1^{\circ}$, and the vertical dotted lines indicate \texttt{log Age} $=8.3$ and 8.5 yr, corresponding to the age thresholds below which the measurement uncertainty for significant number of clusters exceeds $\sigma = 1^{\circ}$ for the 99\% and 95\% enclosed mass fractions, respectively. The uncertainty declines rapidly with age due to the increasing number of escaped particles populating the tails.}
    \label{fig:std_bootstrap}
\end{figure}

Since our simulations contain a finite number of particles, sampling noise can affect the fitted principal graph, particularly at the low-density tail endpoints. To quantify this uncertainty, we bootstrap each distribution 100 times with replacement, refit the principal curve for each bootstrap sample, and compute the resulting distribution of deflection angles.
Fig.~\ref{fig:std_bootstrap} shows the standard deviation of this distribution as a function of age for all simulations, color-coded by their morphological class. The measurement uncertainty rapidly declines with age, falling below $1^{\circ}$ for clusters with \texttt{log Age} $> 8.3$ yr.  We also experimented with lower enclosed mass fractions:  clipping the distribution to 95\% shifts the high-uncertainty region toward older clusters with $\sigma$ falling below $1^{\circ}$ only for clusters with \texttt{log Age} $> 8.5$ yr; it also increases the fraction of cases where the angle measured for the leading and trailing tails are significantly inconsistent. Therefore, we adopt the 99\% enclosed mass fraction as our default. The mean and standard deviation of the bootstrap deflection angle distribution for each cluster and bar model are reported in the published catalog.

\begin{figure}
    \centering
    \includegraphics[width=\columnwidth]{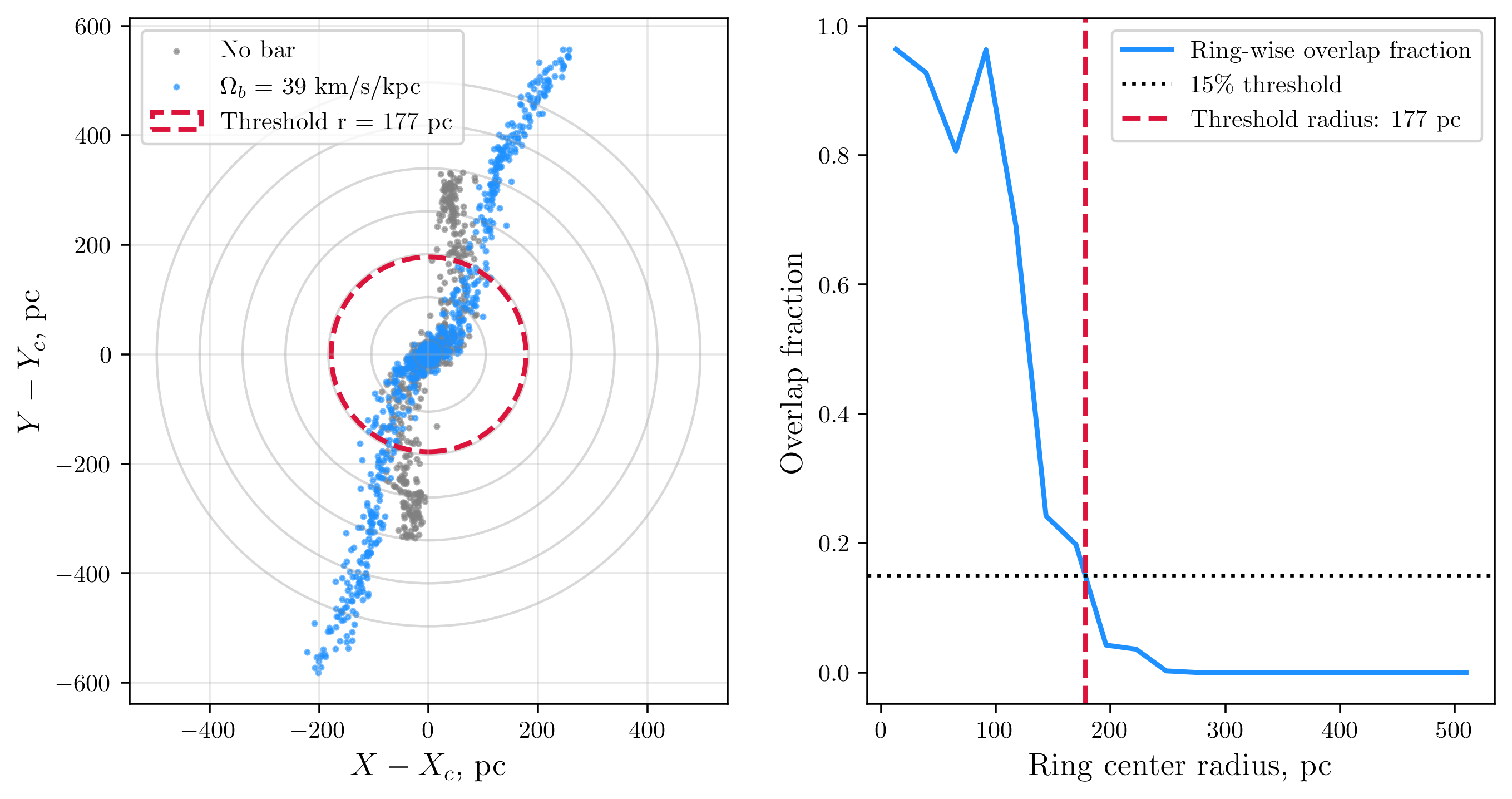}
    \caption{Illustration of the $R_{\mathrm{overlap}}$ calculation. \textit{Left:} particle distributions in the axisymmetric (grey) and barred (blue) models in the cluster-centred $XY$ plane. The dashed red circle marks $R_{\mathrm{overlap}} = 177$ pc, beyond which the two distributions diverge significantly. \textit{Right:} ring-wise overlap fraction as a function of ring centre radius. The horizontal dotted line shows the 15\% threshold, and the vertical dashed line marks $R_{\mathrm{overlap}}$, defined as the largest radius at which the overlap fraction exceeds this threshold.}
    \label{fig:overlap_diagnostic}
\end{figure}

For each simulated cluster in our sample, we determine $R_{\mathrm{overlap}}$, that is  the minimum radial extent required to observe differences in tail orientation arising from different bar pattern speeds. To determine this, we apply the following procedure. We bin the stellar distribution in the $XY$ plane in concentric rings centered on the cluster and compute the convex hull area for particles in each ring for both barred and unbarred models. We then calculate the overlapping fraction of these areas as a function of radius. The overlap radius $R_{\mathrm{overlap}}$ is defined as the largest radius at which the overlapping fraction first drops and remains below 0.15. We use the largest radius rather than the first threshold crossing because the ring-wise overlap fraction might fluctuate significantly in some clusters due to sparse particle sampling in the outer tails. This procedure is illustrated in Figure~\ref{fig:overlap_diagnostic}. The median $R_{\mathrm{overlap}}$ across our sample is 143 pc (16th--84th percentile range: 94--189 pc), and we report this quantity for each cluster in the published catalog.

\subsection{Deflection patterns for the simulated sample}\label{sec:trends}

\begin{figure*}
    \centering
    \includegraphics[width=2\columnwidth]{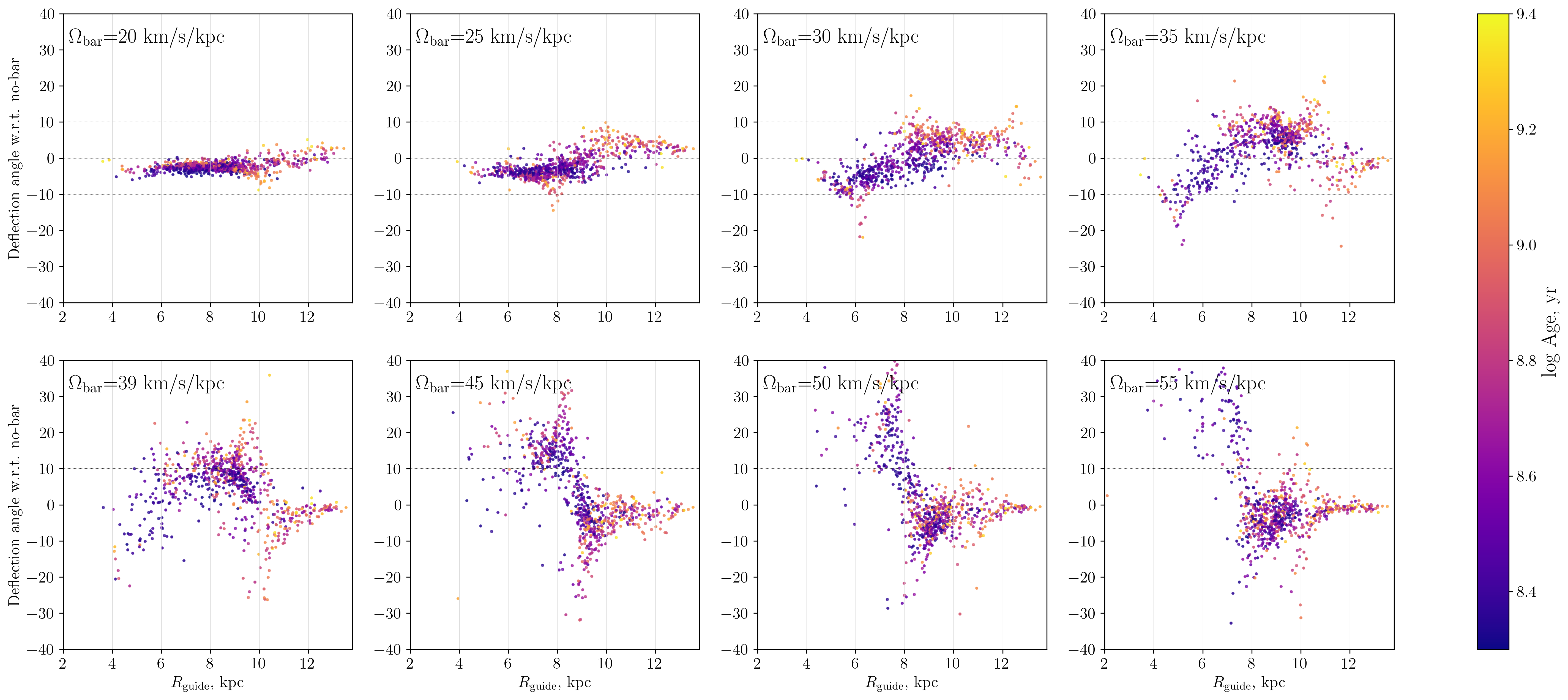}
    \caption{Deflection angle of tidal tails as a function of cluster's galactocentric guiding radius $R_{\mathrm{guide}}$ for different bar pattern speeds $\Omega_b$. Each panel shows results for a barred model with the indicated pattern speed, with points color-coded by the cluster's age. }
    \label{fig:rguide_angle}
\end{figure*}

In this subsection we investigate the trends of the deflection angle with the location of the clusters in the Galaxy. For that analysis we use the mean of the bootstrap distribution as a deflection angle. We limit our sample to clusters with regular shape (morphology class \texttt{0}--\texttt{2}) and with \texttt{log Age} $> 8.3$ yr to exclude cases with uncertain measurements of the deflection, which leaves us with 7188 simulations.

To describe the position of each cluster in the Galaxy, we use its guiding radius $R_{\mathrm{guide}}$ in the axisymmetric potential. We use \texttt{galpy}\footnote{\url{http://github.com/jobovy/galpy}} \citep{galpy} to obtain the circular velocity $v_{\mathrm{circ}}(R)$ of the axisymmetric potential and computed guiding radius from $R_{\mathrm{guide}} v_{\mathrm{circ}}(R_{\mathrm{guide}}) = L_z$, where $L_z$ is the $z-\mathrm{component}$ of angular momentum of the simulated cluster in that potential. 

The distribution of deflection angles as a function of the guiding radius $R_{\mathrm{guide}}$ shows a large variety of behaviours depending on the assumed bar pattern speed, as shown in Fig.~\ref{fig:rguide_angle}, where points are colour-coded by cluster age.
For the slowest bar ($\Omega_b = 20$ km/s/kpc, first row, first subpanel), the majority of clusters exhibit near-zero deflection angles regardless of their galactic position, with some minor deviations beginning to appear in the outer regions. With increasing bar pattern speed, outer clusters begin to show small but predominantly positive deflection angles, while clusters closer to the Galactic centre exhibit more negative deflections (e.g. $\Omega_b = 25$ and 30 km/s/kpc, first row, second and third subpanels). At higher pattern speeds, a clear radial transition appears: clusters just inside a characteristic radius show a strong excess of positive deflections, which sharply reverses to strongly negative deflections just outside it. This transition lies at $R_{\mathrm{guide}} \approx 9-10$ kpc for $\Omega_b = 39$ km/s/kpc and shifts inward with increasing pattern speed reaching $6-8$ kpc for $\Omega_b = 55$ km/s/kpc, while clusters at the larger radii remain mostly unaffected.
In all cases, no systematic trend with cluster age is observed at fixed $R_{\mathrm{guide}}$.

\begin{figure}
    \centering
    \includegraphics[width=\columnwidth]{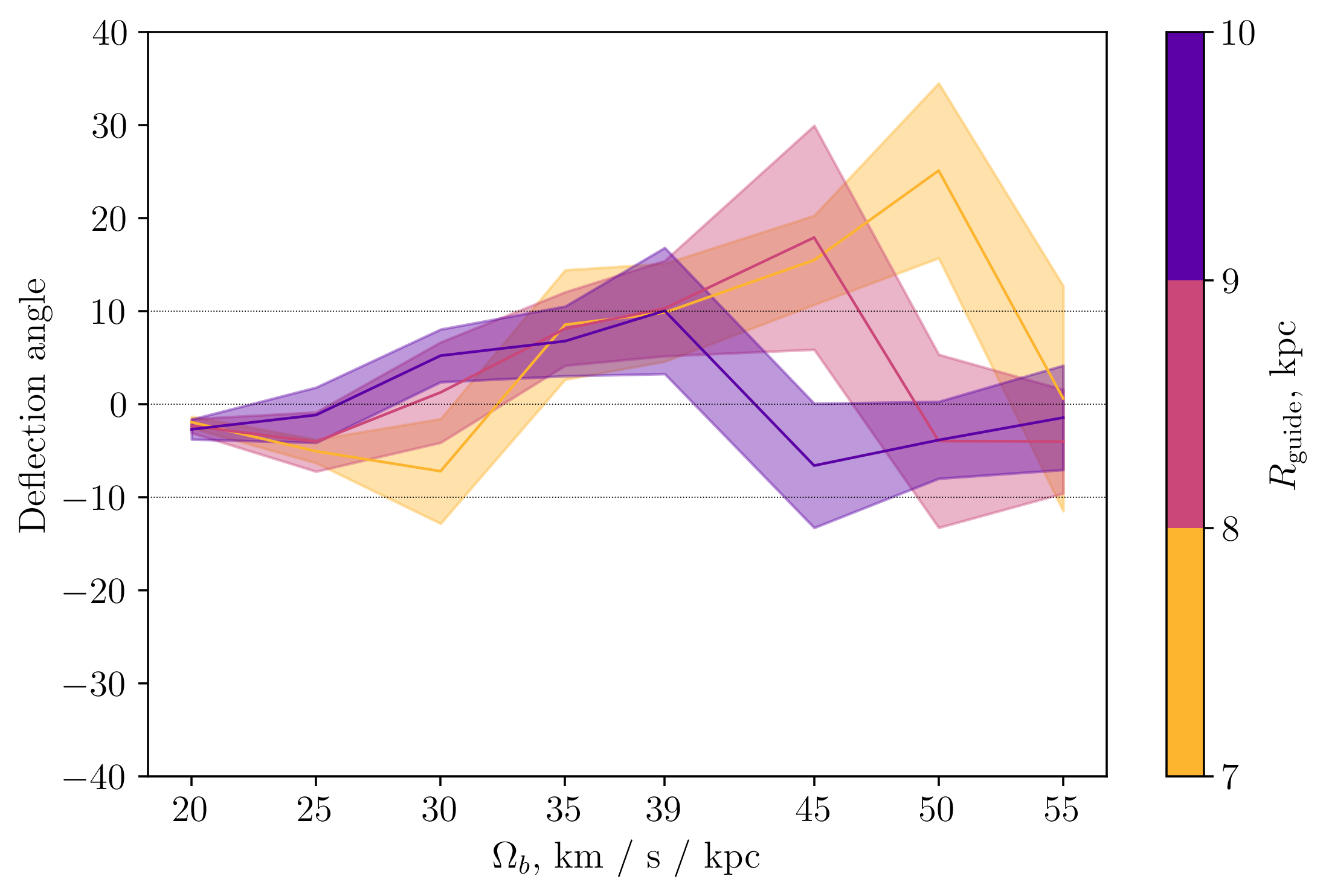}
    \caption{Median deflection angle as a function of bar pattern speed for clusters binned by galactocentric guiding radius $R_{\mathrm{guide}}$. Solid lines show the median deflection angle for each radial bin, with shaded regions indicating $\pm 1\sigma$ scatter. Each radius bin shows the maximum deflection angle at a different $\Omega_b$ -- outermost clusters at moderate, innermost at the highest pattern speeds -- so the deflection profile across guiding radii is characteristic of the bar's pattern speed.}
    \label{fig:shade_omega}
\end{figure}

Fig.~\ref{fig:shade_omega} summarizes the relationship between bar pattern speed, guiding radius, $R_{\mathrm{guide}}$, and deflection angle for our cluster sample. Solid lines show median values of deflection angles for clusters binned by guiding radius $R_{\mathrm{guide}}$ (colour-coded), with shaded regions indicating $\pm 1 \sigma$ scatter. 
Each radius bin follows broadly the same pattern: the tails are largely undeflected at the lowest pattern speeds, then deflection rises to a maximum, declines to a minimum and return back to zero. The pattern is shifted along $\Omega_b$ between bins: the outermost clusters (purple) peak at moderate pattern speeds, while the innermost (yellow) peak at the highest $\Omega_b$. The amplitude varies as well, with the inner clusters the most strongly affected, as they are located closest to the bar.

Because each radial bin is non-monotonic in $\Omega_b$, the deflection of clusters at a single guiding radius does not uniquely determine the pattern speed. The deflection profile \textit{across} guiding radii, however, is characteristic of a given $\Omega_b$: each pattern speed produces a different pattern of which radii are most strongly deflected and which have returned to near-zero. In principle, then, measuring tidal-tail deflection angles for an observed cluster population spanning a range of guiding radii would constrain the Galactic bar's pattern speed, with the constraint tightening as more clusters populate each radius bin.

\subsection{Relation to resonances}\label{sec:resonances}

Fig.~\ref{fig:rguide_angle} and Fig.~\ref{fig:shade_omega} demonstrate that prominent radial features -- in particular, the radial range of maximum deflection angle scatter and the outer boundary beyond which deflections vanish -- shift systematically inward with increasing $\Omega_b$, suggesting a connection to bar resonances whose locations scale inversely with pattern speed. 

In barred potentials, orbits, for which the angular frequency of the orbit relative to the bar $\Omega_{\phi} - \Omega_b$ is commensurate with the frequency of radial oscillations $\Omega_R$, are closed in the bar frame and experience the bar's gravitational perturbation repeatedly at the same orbital phases. This resonant forcing can significantly alter the orbital properties of cluster stars, and hence the orientation of their tidal tails. The most dynamically important resonances are the inner Lindblad resonance (ILR, $\displaystyle \frac{\Omega_{\phi} - \Omega_b}{\Omega_R} = \frac{1}{2}$), corotation (CR, $\displaystyle \frac{\Omega_{\phi} - \Omega_b}{\Omega_R} = 0$), and the outer Lindblad resonance (OLR, $\displaystyle \frac{\Omega_{\phi} - \Omega_b}{\Omega_R} = -\frac{1}{2}$) \citep[][]{BinneyTremaine2008}.
In order to explore the deflection pattern of orbits at resonances and in between them, we use the spectral analysis to obtain angular and radial frequencies for each cluster orbit and evaluate the ratio $\displaystyle \frac{\Omega_{\phi} - \Omega_b}{\Omega_R}$.

\begin{figure}
    \centering
    \includegraphics[width=\columnwidth]{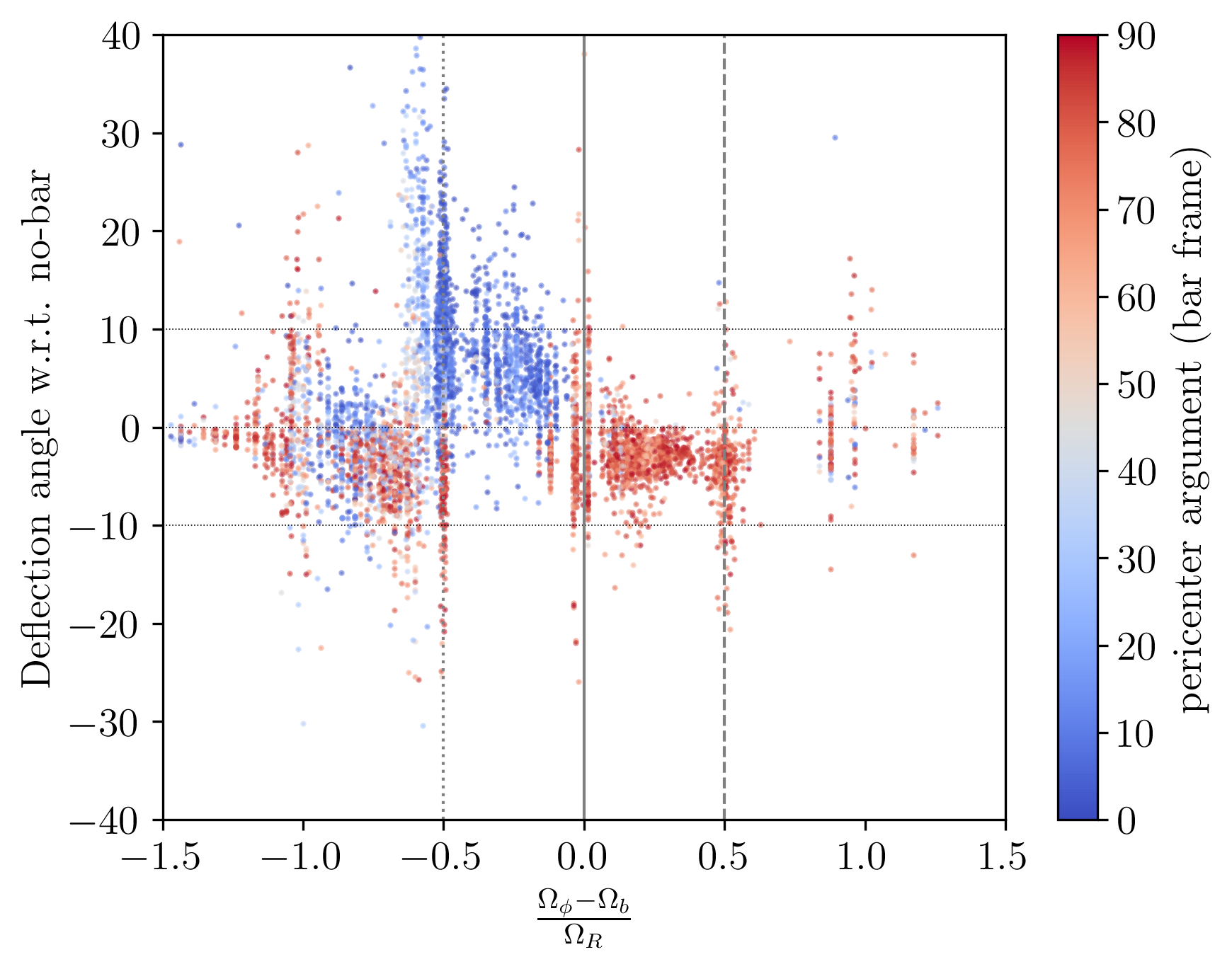}
    \caption{Deflection angle versus frequency ratio $\frac{\Omega_\phi - \Omega_b}{\Omega_R}$ for all simulated clusters, combining data from all eight bar pattern speeds. Points are color-coded by the phase of pericenter $\varpi$ in the bar reference frame (mapped to $[0, \pi/2]$), where blue indicates pericenters aligned with the bar's major axis and red indicates clusters passing pericenters near the minor axis. Vertical lines mark the inner Lindblad resonance (ILR, +0.5), corotation (CR, 0), and outer Lindblad resonance (OLR, -0.5). Clusters near OLR show the largest deflections with a systematic phase dependence: positive angles for major-axis pericenters (blue) and negative angles for minor-axis pericenters (red).}
    \label{fig:freqratio_angle_all_omegas}
\end{figure}

Fig.~\ref{fig:freqratio_angle_all_omegas} shows the deflection angle as a function of the frequency ratio $\displaystyle \frac{\Omega_{\phi} - \Omega_b}{\Omega_R}$ combined for all simulated bar pattern speeds, and highlights that the largest deflections of the tails are observed for orbits near the OLR ($\displaystyle \frac{\Omega_{\phi} - \Omega_b}{\Omega_R} = -1/2$). The colour coding shows the angle between the orbital pericentre (in the bar frame) and the bar's major axis, revealing that the sign of the deflection angle is determined by the alignment of the orbit with the bar.  

For the near-resonant clusters, which exhibit the largest scatter of deflection angles, we examine how tidal tail deflection varies along the orbit. Since clusters in barred and unbarred potentials follow different orbital paths, we cannot directly compare the two tail orientations at the same orbital phase. Instead, we measure the deflection angle of the barred distribution relative to the cluster's instantaneous velocity vector, and subtract the corresponding angle for the unbarred tail -- which is typically small $\sim 3\textup{--}5^{\circ}$ -- to maintain consistency with our original definition of deflection angle.

\begin{figure*}
    \centering
    \includegraphics[width=2\columnwidth]{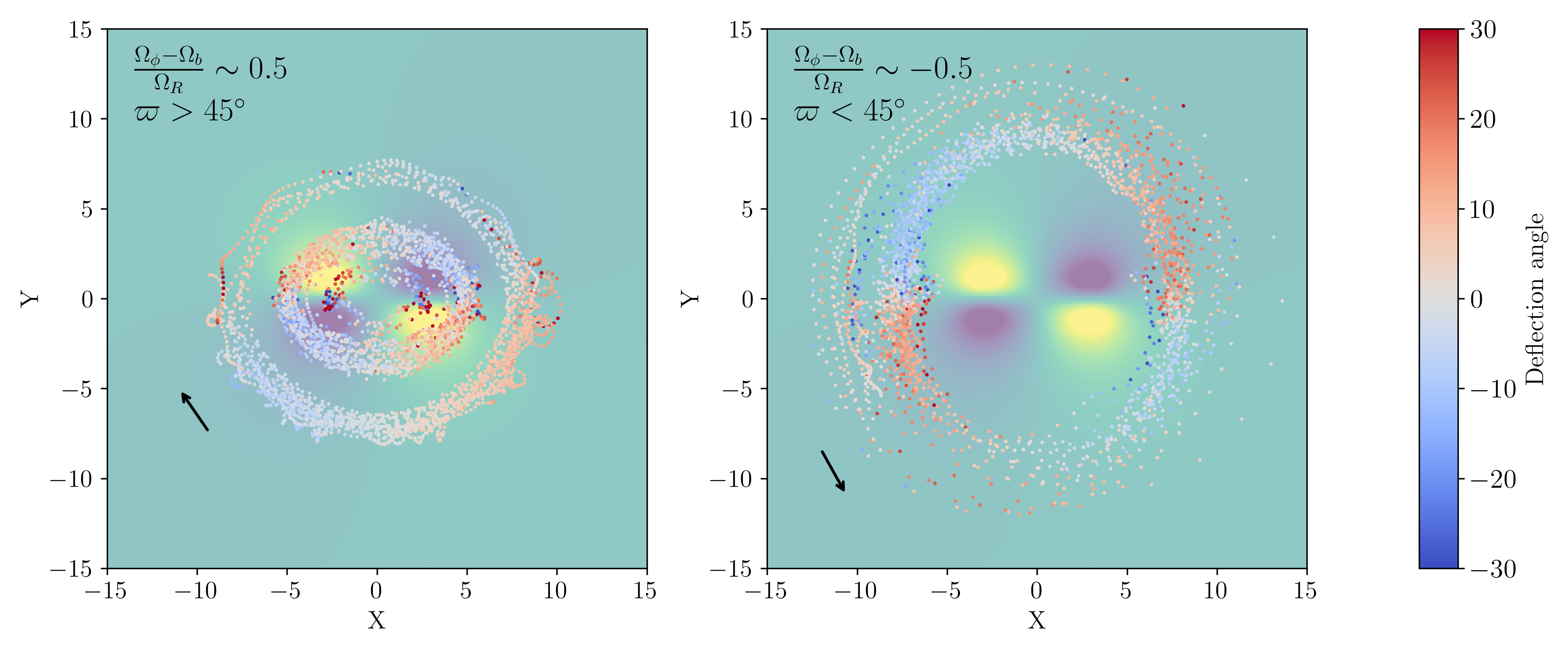}
    \caption{Deflection angle of tidal tails as a function of orbital position in the bar frame, for clusters near the ILR ($\frac{\Omega_\phi - \Omega_b}{\Omega_R} \sim 0.5$, left) and OLR ($\frac{\Omega_\phi - \Omega_b}{\Omega_R} \sim -0.5$, right), filtered by the argument of pericenter relative to the bar major axis ($\varpi > 45^{\circ}$ and $\varpi < 45^{\circ}$, respectively). Points show individual cluster positions along their orbits in the bar frame, colour-coded by deflection angle at that moment. The smooth background shows the torque exerted by the bar potential at each location. The black arrow indicates the direction of revolution for the selected orbits. 
    \LAIA{I wonder if this figure can go to the appendix?}}
    \label{fig:angle_along_orbit}
\end{figure*}

In Fig.~\ref{fig:angle_along_orbit} we show the orbital paths in the bar frame for the clusters near ILR on x1(1) orbits (aligned with the bar) on the left panel and OLR-clusters on x1(2) orbits (perpendicular to the bar) on the right panel. The nomenclature for the orbits families is adopted from \citet{ContopoulosGrosbol1989}. The background is coloured by the value of the torque generated by the bar, and the color of the points represents the deflection angle.
A clear pattern emerges linking deflection angle to orbital position in the bar frame. For OLR x1(2) clusters, positive deflection angles (red colors) dominate in the upper-right and lower-left quadrants (I and III), with the strongest deflections occurring just after pericenter passage, while negative angles dominate near the pericenters but in II and IV quadrants.
In contrast, ILR x1(1) clusters show negative deflections in the I and III quadrants, and positive deflections in II and IV quadrants. 

Due to observational selection, most clusters in our sample are located near the Sun, placing them in the lower-left quadrant of Figure~\ref{fig:angle_along_orbit} at present day. This means that for OLR x1(2) clusters, we should predominantly observe positive deflection angles  and for ILR x1(1) -- mostly negative angles, which is exactly what we see in Fig.~\ref{fig:freqratio_angle_all_omegas}.

\section{Comparison with observational data}\label{sec:observations}
While Fig.~\ref{fig:shade_omega} demonstrates the diagnostic potential of tidal tail deflections for constraining the bar pattern speed, practical implementation of such analysis requires obtaining reliable membership of tidal tail stars extending far from the cluster centre, for a sufficient number of clusters across the Galactic disc, which is an extremely challenging observational task. Since the first detections in the Hyades, tidal tails have been traced for a growing number of nearby clusters with \textit{Gaia} astrometry, using a variety of techniques -- the convergent point method \citep[][]{Roser19a, Roser19b}, compact and self-compact convergent point methods \citep[][]{Jerabkova21, Boffin22, Risbud25}, clustering in various phase-space coordinates \citep[][]{Meingast19, Tarricq22, Meingast21, Bhattacharya22}, self-organizing maps \citep[][]{Pang2022, Tang2019, Zhang2020}, orbital traceback \citep[][]{Vaher23}, and simulation-informed probability methods \citep[][]{Kos24} -- recovering tails from a few tens of parsecs to nearly a kiloparsec in length.

Recently, \citet{Jadhav25} compiled and assessed the reliability of 120 membership catalogues for 58 nearby clusters (within 500 pc of the Sun), assigning quality grades based on morphological criteria and comparison with N-body simulations in axisymmetric potentials. 
We compare a subset of these observed tidal structures with our simulations to investigate whether current data can already provide meaningful constraints on bar parameters, and to identify which clusters should be primary targets for follow-up observations.

\begin{figure*}
    \centering
    \includegraphics[width=2\columnwidth]{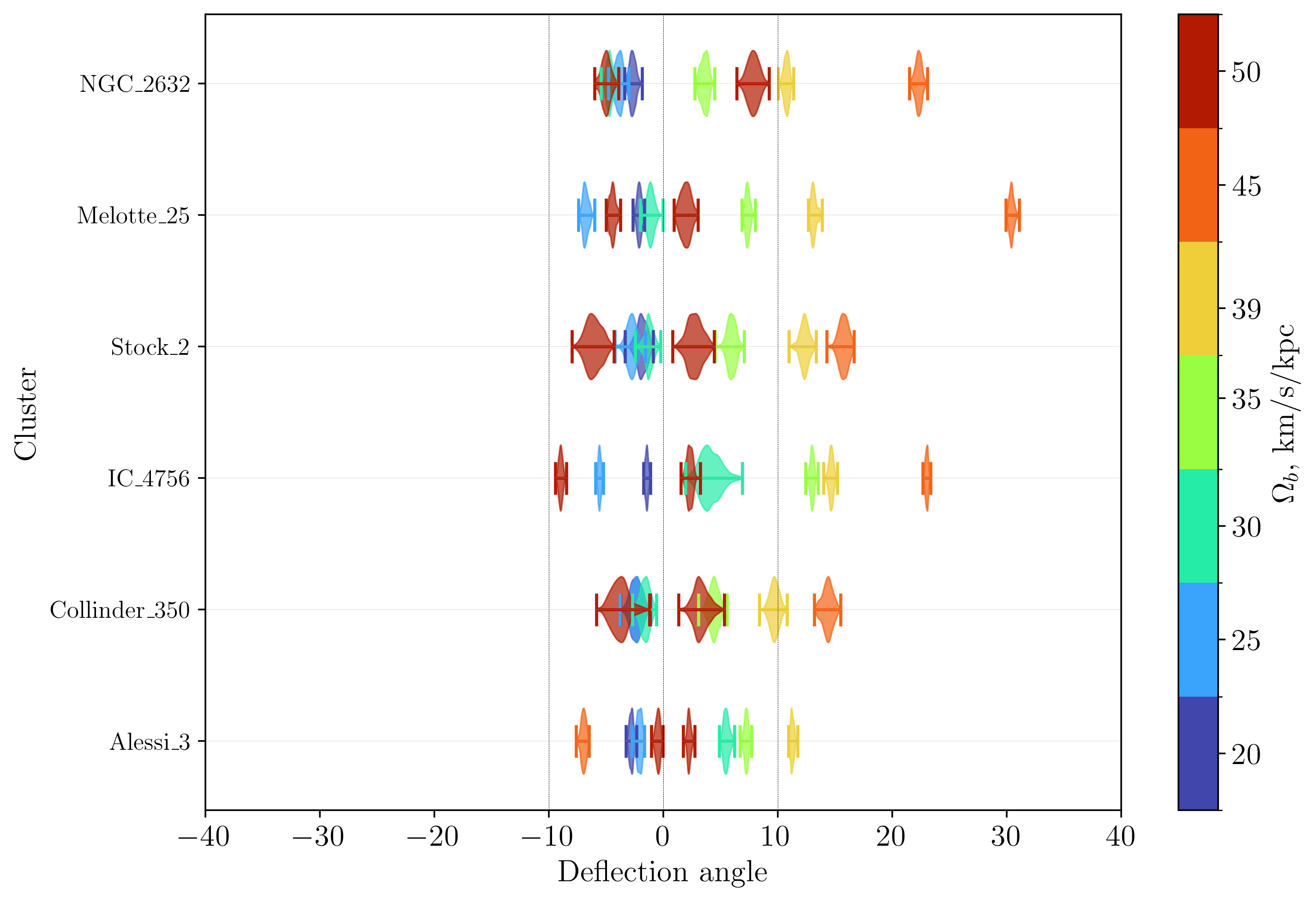}
    \caption{Predicted deflection angles for the six bar-sensitive clusters selected from \citet{Jadhav25}, shown for all eight bar pattern speeds (colour-coded by $\Omega_b$). Each violin shows the distribution of measured deflection angles across bootstrap resamples (see Sect.~\ref{sec:deflection} for details).}
    \label{fig:jadhav_clusters}
\end{figure*}

From the \citet{Jadhav25} sample, we focus on clusters satisfying two criteria: (1) age exceeding the threshold used in the analysis of our simulated sample (\texttt{logAge50 > 8.3}, yielding 22 clusters), and (2) a predicted deflection angle exceeding $10^{\circ}$  for at least one bar pattern speed. This selection yields 6 clusters; two additional clusters satisfying the angle criterion (Ruprecht 147 and NGC 752) are excluded because their large deflections occur only in simulations with morphological class \texttt{3}, for which the deflection angle measurement is unreliable. Figure~\ref{fig:jadhav_clusters} shows the deflection angles from our simulations for the 6 selected clusters across all eight bar pattern speeds, with the bootstrap distribution shown as a violin. To complement the bootstrap uncertainty, which captures sampling noise in the principal curve fit, we also tested the effect of observational uncertainties on the cluster's present-day phase-space coordinates: for each cluster and bar pattern speed we resampled observed astrometry and radial velocity 50 times within its measured uncertainties \citep{HuntReffert24}, transformed to galactocentric initial conditions, and repeated the full simulation and deflection measurement for each realization. The resulting spread is smaller than the bootstrap uncertainty for all clusters and pattern speeds.

\subsection{Case studies}
For each of the 6 selected clusters, 
we visually compare the distribution in the $XY$ plane in galactocentric coordinates from the various catalogues compiled by \citet{Jadhav25} with our simulated tidal tails in both axisymmetric and barred potentials. We convert observed positions and velocities to the galactocentric frame using the solar motion and Galactic parameters listed in Sect.~\ref{sec:data}.

\subsubsection{\textit{NGC 2632}}
\begin{figure*}
    \centering
    \includegraphics[width=2\columnwidth]{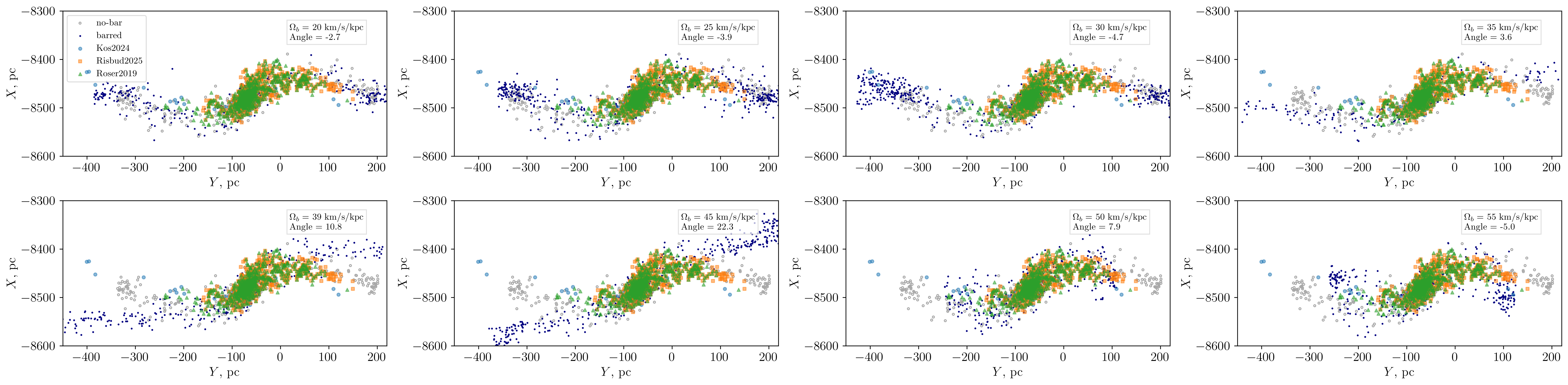}
    \caption{Simulated and observed tidal tails of NGC 2632 (Praesepe) in the Galactocentric $XY$ plane, shown for the eight bar pattern speeds (one per panel). Blue points show the simulated tail in the barred potential at the pattern speed $\Omega_b$ annotated in each panel; grey points show the axisymmetric (no-bar) model. Coloured symbols mark observed members from the literature catalogues indicated in the legend. The deflection angle (in degrees) of the barred model relative to the no-bar case is annotated in each panel.}
    \label{fig:obs_cl_ngc_2632}
\end{figure*}

NGC 2632 (Praesepe, Fig.~\ref{fig:obs_cl_ngc_2632}) presents one of the most informative cases. The \citet{Roser19b} catalogue traces an extended leading tail, identified through the convergent point method. This tail structure agrees well with the results of our axisymmetric model and barred models with slower pattern speed ($\Omega_b \le 30$ km/s/kpc) or fast bars ($\Omega_b \ge 50$ km/s/kpc), and it is clearly inconsistent with the tails predicted by models with moderate $\Omega_b$ (39 and 45 km/s/kpc).

\subsubsection{\textit{Melotte 25 (Hyades)}} 

\begin{figure*}
    \centering
    \includegraphics[width=2\columnwidth]{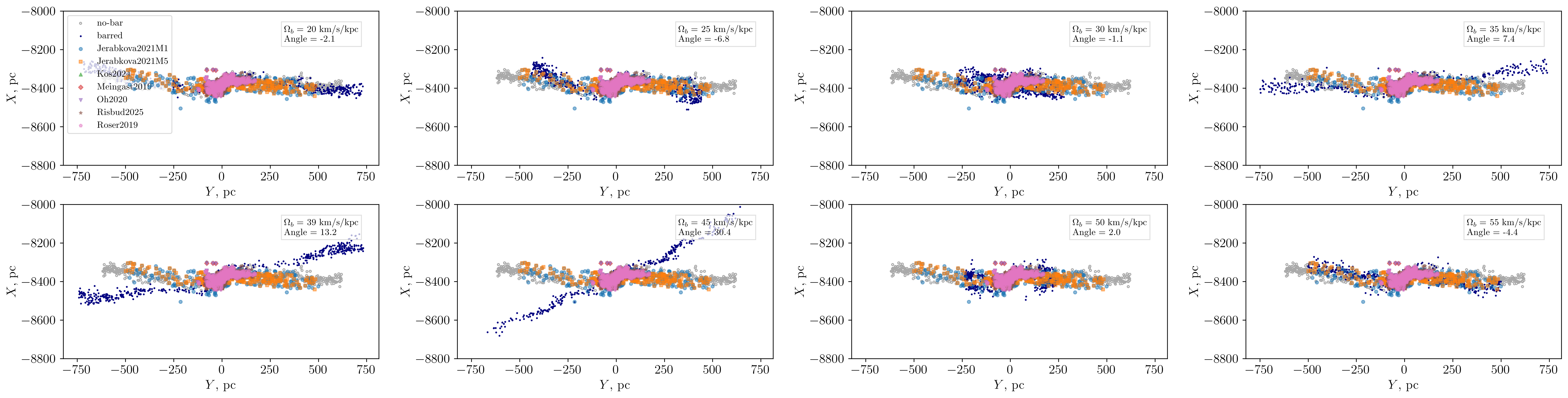}
    \caption{As Fig.~\ref{fig:obs_cl_ngc_2632}, but for Melotte 25 (Hyades).}
    \label{fig:obs_cl_melotte_25}
\end{figure*}

Melotte 25 (Hyades, Fig.~\ref{fig:obs_cl_melotte_25}) is the best-studied case, with the longest tails traced by multiple catalogues \citep{Jerabkova21, Roser19a, Kos24}. The observed tail structures match well the predictions from our axisymmetric model as well as barred models with either the slowest ($\Omega_b \le 25$ km/s/kpc) or fastest ($\Omega_b \ge 50$ km/s/kpc) pattern speeds, where deflections relative to the axisymmetric case remain small. However, the catalogues that provide the most extended tails \citep{Jerabkova21, Kos24} identified the members through comparison with N-body simulations, so their selection might be biased. 
Considering instead the \citet{Roser19a} catalogue, whose membership was determined through the modified convergent point method which is purely data-driven, the Hyades shows the same trend as NGC 2632, consistent with the two clusters lying at similar Galactocentric radii: the majority of bar models produce tails consistent with the observed structure, with the exception of moderate pattern speeds ($\Omega_b = 39$ and 45 km/s/kpc), which predict tail orientations clearly inconsistent with the data. 

\subsubsection{\textit{Stock 2}} 

\begin{figure*}
    \centering
    \includegraphics[width=2\columnwidth]{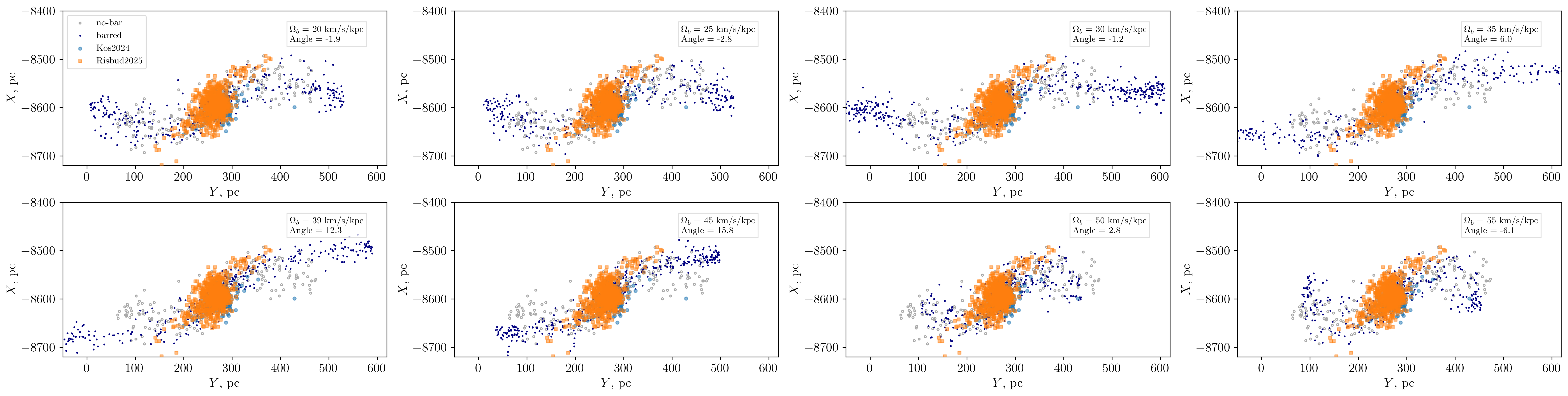}
    \caption{As Fig.~\ref{fig:obs_cl_ngc_2632}, but for Stock 2.}
    \label{fig:obs_cl_stock_2}
\end{figure*}

Stock 2 (Fig.~\ref{fig:obs_cl_stock_2}) provides another interesting case where observed tail stars extend beyond the overlap radius $R_{\mathrm{overlap}}$. The observed tail broadly follows the orientation predicted by the axisymmetric model, while comparison with barred models favours moderate to fast pattern speeds ($\Omega_b \ge 35$ km/s/kpc). However, the limited statistics at the outer radii complicates the comparison. Taken together, NGC 2632, Hyades, and Stock 2 suggests bar pattern speeds in the fast range ($\Omega_b \ge 50$ km/s/kpc), though this constraint remains weak given the sparse members count. 

\subsubsection{\textit{IC 4756}} 

\begin{figure*}
    \centering
    \includegraphics[width=2\columnwidth]{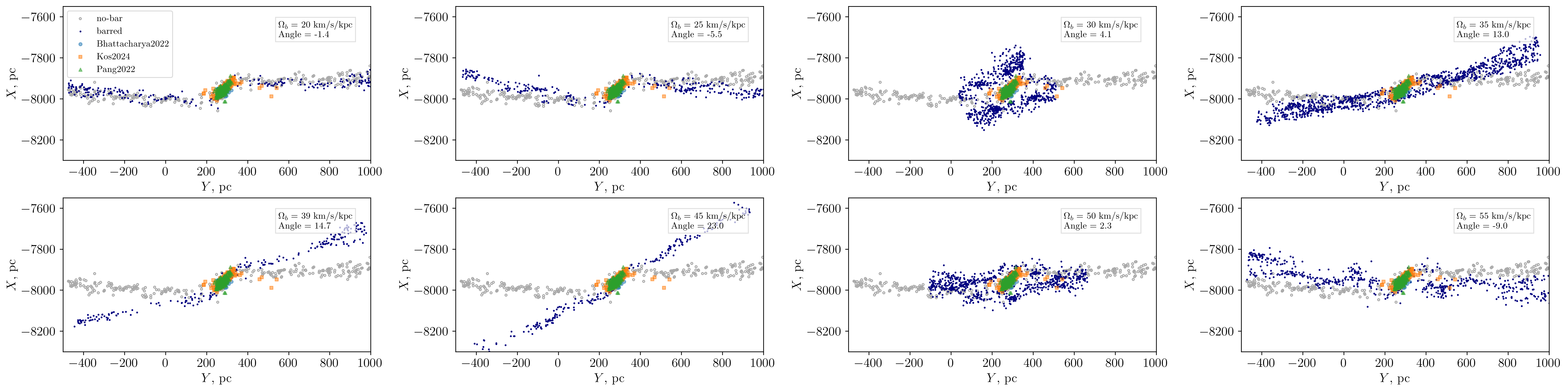}
    \caption{As Fig.~\ref{fig:obs_cl_ngc_2632}, but for IC 4756.}
    \label{fig:obs_cl_ic_4756}
\end{figure*}

IC 4756 (Fig.~\ref{fig:obs_cl_ic_4756}) illustrates both the promise and current limitations of this approach. Our simulations predict distinct morphologies across nearly all pattern speeds, with tails differing not only in orientation but also in their width, making this cluster one of the most discriminative targets for constraining bar properties. Unfortunately, current observations trace only the innermost regions, where predictions from all models converge. The \citet{Kos24} catalogue extends to larger radii and rejects the moderate bar ($\Omega_b = 39$ and 45 km/s/kpc) model prediction; which is unsurprising given that their membership identification method constructs a likelihood function based on comparison with simulations run in fast-bar potential. Nevertheless, IC 4756 is a promising candidate for deeper observations and membership analysis that accounts for bar-induced effects.

\subsubsection{\textit{Collinder 350 and Alessi 3}} 

\begin{figure*}
    \centering
    \includegraphics[width=2\columnwidth]{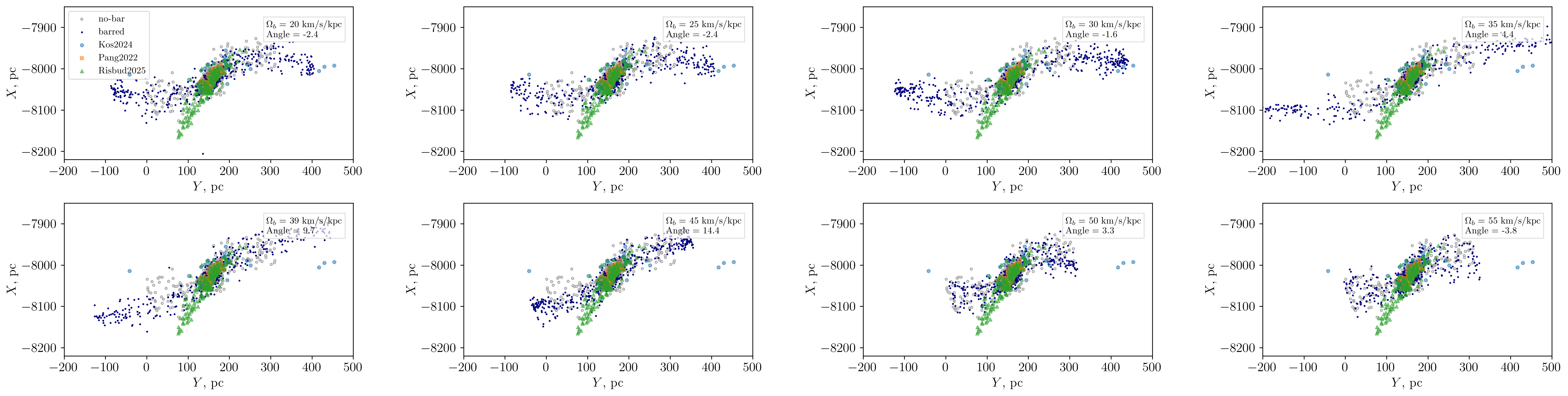}
    \caption{As Fig.~\ref{fig:obs_cl_ngc_2632}, but for Collinder 350.}
    \label{fig:obs_cl_collinder_350}
\end{figure*}

\begin{figure*}
    \centering
    \includegraphics[width=2\columnwidth]{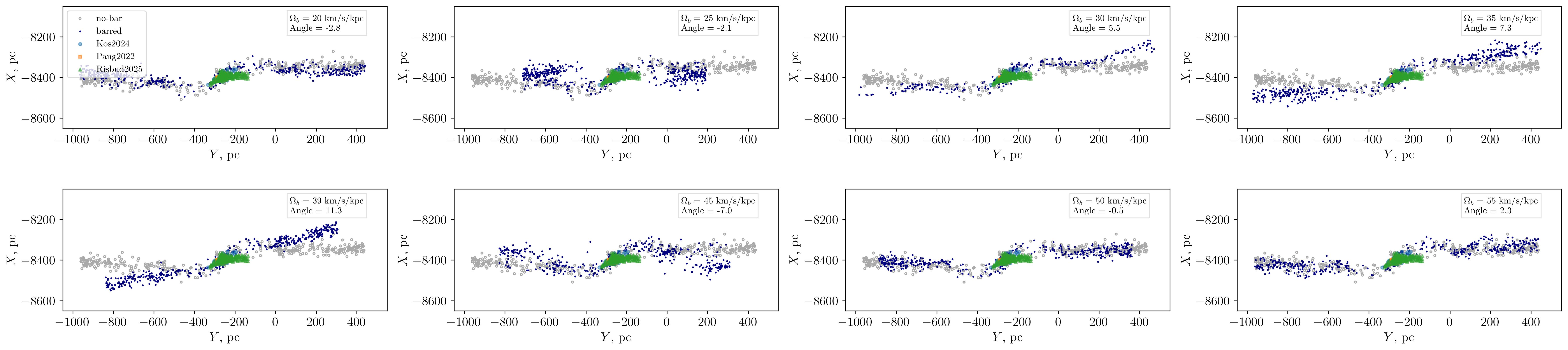}
    \caption{As Fig.~\ref{fig:obs_cl_ngc_2632}, but for Alessi 3.}
    \label{fig:obs_cl_alessi_3}
\end{figure*}

Clusters Collinder 350 and Alessi 3 (Fig.~\ref{fig:obs_cl_collinder_350} and \ref{fig:obs_cl_alessi_3}) present puzzling cases where well-populated tails agree only marginally with any of our simulated tail morphologies. These systematic deviations (in the leading tail for Alessi 3 and trailing for Collinder 350 seen in data from \citet{Risbud25} catalogue marked with green triangles) suggest either that the membership obtained for these clusters with self-compact convergent point method is unreliable, or that these clusters have experienced perturbations beyond the smooth bar and disc potential -- such as encounters with giant molecular clouds -- that are not captured by our simulations.

\subsection{Implications for observational catalogues}

To provide practical guidance for future observational studies, we classify clusters in our sample according to their sensitivity to the bar potential.
For each cluster, we compute the maximum absolute deflection angle across all eight bar pattern speeds. Clusters with maximum absolute deflections below $10^{\circ}$ are labelled ``bar-insensitive''; for these systems, barred and unbarred models predict nearly identical tail orientations. In contrast, clusters with maximum deflections exceeding $10^{\circ}$ are ``bar-sensitive'': their tail orientation is altered by the bar, making them the most promising candidates for constraining the pattern speed, but also requiring careful treatment in membership analysis. 

This distinction has direct implications for interpreting existing tidal tail catalogues. For example, \citet{Kos24} identified tidal tail members for 476 open clusters using likelihoods constructed from N-body simulations. Of the 354 clusters that overlap with our sample, 152 are bar-insensitive, so their membership is robust to the choice of potential. The membership assignments for the remaining clusters may be biased by the specific bar model adopted. 

\section{Discussion}\label{sec:discussion}
Our study highlights the importance of accounting for bar perturbations when analysing the morphology of open cluster tidal tails -- a timely contribution given the growing number of works focused on identifying members beyond the tidal radius. The problem is two-fold. On the one hand, catalogues recovering the most extended tails are often constructed with guidance from N-body simulations, so the recovered morphologies may be biased by the choice of underlying Galactic potential, which is often assumed to be axisymmetric. On the other hand, the potential of open clusters as probes of the Galactic bar and spiral arms is increasingly recognised, with recent studies \citep[e.g.,][]{Thomas2023, Zhou2026} using the Hyades to constrain the pattern speed of the non-axisymmetric structures -- making it important to understand which clusters are most sensitive to bar perturbations, and in what way. However, \citet{Thomas2023} demonstrated that field contamination in the Galactic disc is strong enough that simulation-informed membership selections constructed with different bar models all recover comparable numbers of candidates, so current data cannot meaningfully discriminate between bar pattern speeds from the Hyades alone. Overcoming this degeneracy will require a larger sample of bar-sensitive clusters analysed jointly and additional discriminating information -- for instance from spectroscopy or precise stellar ages -- to separate true members from disc contaminants. The sensitivity classification we provide here is a step toward identifying that sample.

While the analysis of simulated tail orientations allows us to classify clusters by their sensitivity to the bar, our results are subject to several caveats. First, we adopt uniform initial masses for all clusters; while the orbits of escaped particles are governed by the Galactic potential alone and are therefore largely independent of cluster mass, the initial mass does affect escape velocities and consequently tail length. This means that $R_{\mathrm{overlap}}$, which marks the radius at which differences between bar models become apparent, may in principle scale with tail length and hence with cluster mass -- a dependence we leave for future investigation. 
With more careful treatment of tail length, it could also serve as a complementary bar diagnostic: \citet{Hattori2016} showed that the growth rate of tidal streams in barred potentials depends on the alignment of pericentric passages with the bar's major and minor axes, meaning that the bar can produce streams that are substantially shorter or longer than expected from axisymmetric models -- an effect we also qualitatively observe in our simulations.
The stellar counts along the tails are also not realistic, owing to the uniform number of particles assigned for each cluster, the non-collisional nature of the simulations, and the absence of stellar evolution -- all of which would affect the number of stars populating the tails in practice. 

On the side of the Galactic potential, we adopt several additional simplifications. We fix the bar length to the same value across all pattern speeds, whereas in reality the bar length is not precisely constrained. We do not include spiral arms, which can also induce tail deflections; however, \citet{Zhou2026} find that spiral arms have a lesser effect than the bar, at least for the Hyades. The choice of axisymmetric background potential introduces additional uncertainty: for example, \citet{Jerabkova21} show that different axisymmetric Galactic potential models -- specifically the \citet{AllenSantillan} and \citet{Irrgang2013} potentials -- shift the location of epicyclic overdensities in the Hyades tails. Finally, we assume a constant bar pattern speed, which is reasonable for young clusters but less so for older ones: for a decelerating bar the resonance location is shifting, and a cluster currently at the resonance may have had a more complex orbital history than our simulations capture.

Additionally, in reality open clusters may encounter other non-axisymmetric perturbations -- such as giant molecular clouds or satellite galaxies -- which can significantly alter the orbital properties of the cluster \citep{Wiggins25} and the morphology of its tidal tails \citep{Jerabkova21, Miller2025}. Such encounters are stochastical: though their statistical effect on cluster populations can be modelled, the specific encounter history of an individual cluster is generally unknown. This may explain the cases where observed tail orientations agree poorly with all of our simulated models, such as Collinder 350 and Alessi 3.

\section{Conclusion}\label{sec:conclusion}
We have carried out a systematic study of how the Galactic bar affects the tidal tails of open clusters, simulating 1453 clusters from the \citet{HuntReffert24} catalogue in an axisymmetric potential and in eight barred potentials with pattern speeds $\Omega_b = 20\textup{--}55$ km/s/kpc. Our main findings are as follows.

The presence of the bar produces a great diversity of tail morphologies, affect the length of the tails, and induce a misalignment of the tails relative to the axisymmetric case. We quantified the orientation difference between barred and unbarred models through the deflection angle, measured from principal curves fitted to the simulated particle distributions.

The deflection angle varies systematically with bar pattern speed and cluster guiding radius. At low pattern speeds the tails remain largely undeflected at all radii; as $\Omega_b$ increases, each guiding-radius bin develops a deflection that rises to a maximum, drops to a minimum, and then returns toward zero. The whole pattern shifts to higher $\Omega_b$ for smaller guiding radii and reaches its largest amplitude for the innermost clusters, whose response peaks near the upper end of our sampled pattern speeds and does not complete the cycle within the simulated range. This migration reflects the inward shift of the bar's resonant radii with increasing pattern speed. Because each radius bin responds non-monotonically to the change of $\Omega_b$, the deflection profile across guiding radii -- rather than any single cluster -- encodes the pattern speed, and could in principle be used to constrain it from an observed cluster population.
The strongest deflections occur for clusters whose orbital frequencies place them near the outer Lindblad resonance, with the sign of the deflection determined by the angle between the orbital pericentre and the bar's major axis. 

We defined an overlap radius $R_{\mathrm{overlap}}$ as the distance from the cluster centre beyond which the tail orientations of different bar models become distinguishable, and found a median value of 143 pc across the sample. We further classified each cluster as bar-sensitive or bar-insensitive based on whether its maximum absolute deflection across the eight bar models exceeds $10^{\circ}$. 

Comparing our simulations to observational data compiled by \citet{Jadhav25} for six bar-sensitive clusters, we found that the tails of NGC 2632 (Praesepe) and Melotte 25 (Hyades) are inconsistent with moderate pattern speeds ($\Omega_b \sim 39\textup{--}45$ km/s/kpc) and broadly favour either slow or fast bars. The constraint is weak in isolation, but it illustrates the method's potential, which will be realized as reliable tail membership beyond $R_{\mathrm{overlap}}$ becomes available for a larger sample of bar-sensitive clusters.

The catalogue of deflection angles, morphological classes, overlap radii, and bar-sensitivity flags we provide is intended to serve both as guidance for future observational searches and as a tool for re-assessing existing tidal tail catalogues. Tightening the constraint on the bar pattern speed from open cluster tails will require deeper membership identification extending beyond $R_{\mathrm{overlap}}$ for a statistically larger sample of bar-sensitive clusters, ideally combined with complementary information such as spectroscopy or precise ages to reduce disc contamination.

\section*{Acknowledgements}
HP and SF acknowledge the support from the Centre national d’études spatiales (CNES) through a postdoctoral fellowship.
 This work has made use of the computational resources available at the Paris Observatory

\section*{Data Availability}

The catalogue of deflection angles, morphological classes, overlap radii $R_{\mathrm{overlap}}$, and bar-sensitivity flags, together with the simulation code, the underlying dataset, and an interactive viewer of the simulated particle distributions, will be made publicly available upon acceptance of this article.



\bibliographystyle{mnras}
\bibliography{example} 

@ARTICLE{HuntReffert24,
       author = {{Hunt}, Emily L. and {Reffert}, Sabine},
        title = "{Improving the open cluster census. III. Using cluster masses, radii, and dynamics to create a cleaned open cluster catalogue}",
      journal = {\aap},
         year = 2024,
        month = jun,
       volume = {686},
          eid = {A42},
        pages = {A42},
          doi = {10.1051/0004-6361/202348662},
archivePrefix = {arXiv},
       eprint = {2403.05143},
 primaryClass = {astro-ph.GA},
       adsurl = {https://ui.adsabs.harvard.edu/abs/2024A&A...686A..42H}
}

@ARTICLE{MiyamotoNagai1975,
       author = {{Miyamoto}, M. and {Nagai}, R.},
        title = "{Three-Dimensional Models for the Distribution of Mass in Galaxies}",
      journal = {\pasj},
         year = 1975,
        month = dec,
       volume = {27},
       number = {4},
        pages = {533-543},
          doi = {10.1093/pasj/27.4.533},
       adsurl = {https://ui.adsabs.harvard.edu/abs/1975PASJ...27..533M}
}

@ARTICLE{Khalil25,
       author = {{Khalil}, Y.~R. and {Famaey}, B. and {Monari}, G. and {Bernet}, M. and {Siebert}, A. and {Ibata}, R. and {Thomas}, G.~F. and {Ramos}, P. and {Antoja}, T. and {Li}, C. and {Rozier}, S. and {Romero-G{\'o}mez}, M.},
        title = "{A non-axisymmetric potential for the Milky Way disk}",
      journal = {\aap},
         year = 2025,
        month = jul,
       volume = {699},
          eid = {A263},
        pages = {A263},
          doi = {10.1051/0004-6361/202453077},
archivePrefix = {arXiv},
       eprint = {2411.12800},
 primaryClass = {astro-ph.GA},
       adsurl = {https://ui.adsabs.harvard.edu/abs/2025A&A...699A.263K}
}

@ARTICLE{Hattori2016,
       author = {{Hattori}, Kohei and {Erkal}, Denis and {Sanders}, Jason L.},
        title = "{Shepherding tidal debris with the Galactic bar: the Ophiuchus stream}",
      journal = {\mnras},
         year = 2016,
        month = jul,
       volume = {460},
       number = {1},
        pages = {497-512},
          doi = {10.1093/mnras/stw1006},
archivePrefix = {arXiv},
       eprint = {1512.04536},
 primaryClass = {astro-ph.GA},
       adsurl = {https://ui.adsabs.harvard.edu/abs/2016MNRAS.460..497H}
}

@ARTICLE{Pouliasis17,
       author = {{Pouliasis}, E. and {Di Matteo}, P. and {Haywood}, M.},
        title = "{A Milky Way with a massive, centrally concentrated thick disc: new Galactic mass models for orbit computations}",
      journal = {\aap},
         year = 2017,
        month = feb,
       volume = {598},
          eid = {A66},
        pages = {A66},
          doi = {10.1051/0004-6361/201527346},
archivePrefix = {arXiv},
       eprint = {1611.07979},
 primaryClass = {astro-ph.GA},
       adsurl = {https://ui.adsabs.harvard.edu/abs/2017A&A...598A..66P}
}

@ARTICLE{Malhan2018,
       author = {{Malhan}, Khyati and {Ibata}, Rodrigo A.},
        title = "{STREAMFINDER - I. A new algorithm for detecting stellar streams}",
      journal = {\mnras},
         year = 2018,
        month = jul,
       volume = {477},
       number = {3},
        pages = {4063-4076},
          doi = {10.1093/mnras/sty912},
archivePrefix = {arXiv},
       eprint = {1804.11338},
 primaryClass = {astro-ph.GA},
       adsurl = {https://ui.adsabs.harvard.edu/abs/2018MNRAS.477.4063M}
}

@ARTICLE{Shipp2018,
       author = {{Shipp}, N. and {Drlica-Wagner}, A. and {Balbinot}, E. and {Ferguson}, P. and {Erkal}, D. and {Li}, T.~S. and {Bechtol}, K. and {Belokurov}, V. and {Buncher}, B. and {Carollo}, D. and {Carrasco Kind}, M. and {Kuehn}, K. and {Marshall}, J.~L. and {Pace}, A.~B. and {Rykoff}, E.~S. and {Sevilla-Noarbe}, I. and {Sheldon}, E. and {Strigari}, L. and {Vivas}, A.~K. and {Yanny}, B. and {Zenteno}, A. and {Abbott}, T.~M.~C. and {Abdalla}, F.~B. and {Allam}, S. and {Avila}, S. and {Bertin}, E. and {Brooks}, D. and {Burke}, D.~L. and {Carretero}, J. and {Castander}, F.~J. and {Cawthon}, R. and {Crocce}, M. and {Cunha}, C.~E. and {D'Andrea}, C.~B. and {da Costa}, L.~N. and {Davis}, C. and {De Vicente}, J. and {Desai}, S. and {Diehl}, H.~T. and {Doel}, P. and {Evrard}, A.~E. and {Flaugher}, B. and {Fosalba}, P. and {Frieman}, J. and {Garc{\'\i}a-Bellido}, J. and {Gaztanaga}, E. and {Gerdes}, D.~W. and {Gruen}, D. and {Gruendl}, R.~A. and {Gschwend}, J. and {Gutierrez}, G. and {Hartley}, W. and {Honscheid}, K. and {Hoyle}, B. and {James}, D.~J. and {Johnson}, M.~D. and {Krause}, E. and {Kuropatkin}, N. and {Lahav}, O. and {Lin}, H. and {Maia}, M.~A.~G. and {March}, M. and {Martini}, P. and {Menanteau}, F. and {Miller}, C.~J. and {Miquel}, R. and {Nichol}, R.~C. and {Plazas}, A.~A. and {Romer}, A.~K. and {Sako}, M. and {Sanchez}, E. and {Santiago}, B. and {Scarpine}, V. and {Schindler}, R. and {Schubnell}, M. and {Smith}, M. and {Smith}, R.~C. and {Sobreira}, F. and {Suchyta}, E. and {Swanson}, M.~E.~C. and {Tarle}, G. and {Thomas}, D. and {Tucker}, D.~L. and {Walker}, A.~R. and {Wechsler}, R.~H. and {DES Collaboration}},
        title = "{Stellar Streams Discovered in the Dark Energy Survey}",
      journal = {\apj},
         year = 2018,
        month = aug,
       volume = {862},
       number = {2},
          eid = {114},
        pages = {114},
          doi = {10.3847/1538-4357/aacdab},
archivePrefix = {arXiv},
       eprint = {1801.03097},
 primaryClass = {astro-ph.GA},
       adsurl = {https://ui.adsabs.harvard.edu/abs/2018ApJ...862..114S}
}

@ARTICLE{Grillmair2006,
       author = {{Grillmair}, C.~J. and {Dionatos}, O.},
        title = "{Detection of a 63{\textdegree} Cold Stellar Stream in the Sloan Digital Sky Survey}",
      journal = {\apjl},
         year = 2006,
        month = may,
       volume = {643},
       number = {1},
        pages = {L17-L20},
          doi = {10.1086/505111},
archivePrefix = {arXiv},
       eprint = {astro-ph/0604332},
 primaryClass = {astro-ph},
       adsurl = {https://ui.adsabs.harvard.edu/abs/2006ApJ...643L..17G}
}

@ARTICLE{GaiaDR2,
       author = {{Gaia Collaboration} and {Brown}, A.~G.~A. and {Vallenari}, A. and {Prusti}, T. and {de Bruijne}, J.~H.~J. and {Babusiaux}, C. and {Bailer-Jones}, C.~A.~L. and {Biermann}, M. and {Evans}, D.~W. and {Eyer}, L. and {Jansen}, F. and {Jordi}, C. and {Klioner}, S.~A. and {Lammers}, U. and {Lindegren}, L. and {Luri}, X. and {Mignard}, F. and {Panem}, C. and {Pourbaix}, D. and {Randich}, S. and {Sartoretti}, P. and {Siddiqui}, H.~I. and {Soubiran}, C. and {van Leeuwen}, F. and {Walton}, N.~A. and {Arenou}, F. and {Bastian}, U. and {Cropper}, M. and {Drimmel}, R. and {Katz}, D. and {Lattanzi}, M.~G. and {Bakker}, J. and {Cacciari}, C. and {Casta{\~n}eda}, J. and {Chaoul}, L. and {Cheek}, N. and {De Angeli}, F. and {Fabricius}, C. and {Guerra}, R. and {Holl}, B. and {Masana}, E. and {Messineo}, R. and {Mowlavi}, N. and {Nienartowicz}, K. and {Panuzzo}, P. and {Portell}, J. and {Riello}, M. and {Seabroke}, G.~M. and {Tanga}, P. and {Th{\'e}venin}, F. and {Gracia-Abril}, G. and {Comoretto}, G. and {Garcia-Reinaldos}, M. and {Teyssier}, D. and {Altmann}, M. and {Andrae}, R. and {Audard}, M. and {Bellas-Velidis}, I. and {Benson}, K. and {Berthier}, J. and {Blomme}, R. and {Burgess}, P. and {Busso}, G. and {Carry}, B. and {Cellino}, A. and {Clementini}, G. and {Clotet}, M. and {Creevey}, O. and {Davidson}, M. and {De Ridder}, J. and {Delchambre}, L. and {Dell'Oro}, A. and {Ducourant}, C. and {Fern{\'a}ndez-Hern{\'a}ndez}, J. and {Fouesneau}, M. and {Fr{\'e}mat}, Y. and {Galluccio}, L. and {Garc{\'\i}a-Torres}, M. and {Gonz{\'a}lez-N{\'u}{\~n}ez}, J. and {Gonz{\'a}lez-Vidal}, J.~J. and {Gosset}, E. and {Guy}, L.~P. and {Halbwachs}, J.-L. and {Hambly}, N.~C. and {Harrison}, D.~L. and {Hern{\'a}ndez}, J. and {Hestroffer}, D. and {Hodgkin}, S.~T. and {Hutton}, A. and {Jasniewicz}, G. and {Jean-Antoine-Piccolo}, A. and {Jordan}, S. and {Korn}, A.~J. and {Krone-Martins}, A. and {Lanzafame}, A.~C. and {Lebzelter}, T. and {L{\"o}ffler}, W. and {Manteiga}, M. and {Marrese}, P.~M. and {Mart{\'\i}n-Fleitas}, J.~M. and {Moitinho}, A. and {Mora}, A. and {Muinonen}, K. and {Osinde}, J. and {Pancino}, E. and {Pauwels}, T. and {Petit}, J.-M. and {Recio-Blanco}, A. and {Richards}, P.~J. and {Rimoldini}, L. and {Robin}, A.~C. and {Sarro}, L.~M. and {Siopis}, C. and {Smith}, M. and {Sozzetti}, A. and {S{\"u}veges}, M. and {Torra}, J. and {van Reeven}, W. and {Abbas}, U. and {Abreu Aramburu}, A. and {Accart}, S. and {Aerts}, C. and {Altavilla}, G. and {{\'A}lvarez}, M.~A. and {Alvarez}, R. and {Alves}, J. and {Anderson}, R.~I. and {Andrei}, A.~H. and {Anglada Varela}, E. and {Antiche}, E. and {Antoja}, T. and {Arcay}, B. and {Astraatmadja}, T.~L. and {Bach}, N. and {Baker}, S.~G. and {Balaguer-N{\'u}{\~n}ez}, L. and {Balm}, P. and {Barache}, C. and {Barata}, C. and {Barbato}, D. and {Barblan}, F. and {Barklem}, P.~S. and {Barrado}, D. and {Barros}, M. and {Barstow}, M.~A. and {Bartholom{\'e} Mu{\~n}oz}, S. and {Bassilana}, J.-L. and {Becciani}, U. and {Bellazzini}, M. and {Berihuete}, A. and {Bertone}, S. and {Bianchi}, L. and {Bienaym{\'e}}, O. and {Blanco-Cuaresma}, S. and {Boch}, T. and {Boeche}, C. and {Bombrun}, A. and {Borrachero}, R. and {Bossini}, D. and {Bouquillon}, S. and {Bourda}, G. and {Bragaglia}, A. and {Bramante}, L. and {Breddels}, M.~A. and {Bressan}, A. and {Brouillet}, N. and {Br{\"u}semeister}, T. and {Brugaletta}, E. and {Bucciarelli}, B. and {Burlacu}, A. and {Busonero}, D. and {Butkevich}, A.~G. and {Buzzi}, R. and {Caffau}, E. and {Cancelliere}, R. and {Cannizzaro}, G. and {Cantat-Gaudin}, T. and {Carballo}, R. and {Carlucci}, T. and {Carrasco}, J.~M. and {Casamiquela}, L. and {Castellani}, M. and {Castro-Ginard}, A. and {Charlot}, P. and {Chemin}, L. and {Chiavassa}, A. and {Cocozza}, G. and {Costigan}, G. and {Cowell}, S. and {Crifo}, F. and {Crosta}, M. and {Crowley}, C. and {Cuypers}, J. and {Dafonte}, C. and {Damerdji}, Y. and {Dapergolas}, A. and {David}, P. and {David}, M. and {de Laverny}, P. and {De Luise}, F.},
        title = "{Gaia Data Release 2. Summary of the contents and survey properties}",
      journal = {\aap},
         year = 2018,
        month = aug,
       volume = {616},
          eid = {A1},
        pages = {A1},
          doi = {10.1051/0004-6361/201833051},
archivePrefix = {arXiv},
       eprint = {1804.09365},
 primaryClass = {astro-ph.GA},
       adsurl = {https://ui.adsabs.harvard.edu/abs/2018A&A...616A...1G}
}

@ARTICLE{Odenkirchen2001,
       author = {{Odenkirchen}, Michael and {Grebel}, Eva K. and {Rockosi}, Constance M. and {Dehnen}, Walter and {Ibata}, Rodrigo and {Rix}, Hans-Walter and {Stolte}, Andrea and {Wolf}, Christian and {Anderson}, Jr., John E. and {Bahcall}, Neta A. and {Brinkmann}, Jon and {Csabai}, Istv{\'a}n and {Hennessy}, G. and {Hindsley}, Robert B. and {Ivezi{\'c}}, {\v{Z}}eljko and {Lupton}, Robert H. and {Munn}, Jeffrey A. and {Pier}, Jeffrey R. and {Stoughton}, Chris and {York}, Donald G.},
        title = "{Detection of Massive Tidal Tails around the Globular Cluster Palomar 5 with Sloan Digital Sky Survey Commissioning Data}",
      journal = {\apjl},
         year = 2001,
        month = feb,
       volume = {548},
       number = {2},
        pages = {L165-L169},
          doi = {10.1086/319095},
archivePrefix = {arXiv},
       eprint = {astro-ph/0012311},
 primaryClass = {astro-ph},
       adsurl = {https://ui.adsabs.harvard.edu/abs/2001ApJ...548L.165O}
}

@ARTICLE{BaumgardtKroupa2007,
       author = {{Baumgardt}, H. and {Kroupa}, P.},
        title = "{A comprehensive set of simulations studying the influence of gas expulsion on star cluster evolution}",
      journal = {\mnras},
         year = 2007,
        month = oct,
       volume = {380},
       number = {4},
        pages = {1589-1598},
          doi = {10.1111/j.1365-2966.2007.12209.x},
archivePrefix = {arXiv},
       eprint = {0707.1944},
 primaryClass = {astro-ph},
       adsurl = {https://ui.adsabs.harvard.edu/abs/2007MNRAS.380.1589B}
}

@ARTICLE{Ramos2018,
       author = {{Ramos}, P. and {Antoja}, T. and {Figueras}, F.},
        title = "{Riding the kinematic waves in the Milky Way disk with Gaia}",
      journal = {\aap},
         year = 2018,
        month = nov,
       volume = {619},
          eid = {A72},
        pages = {A72},
          doi = {10.1051/0004-6361/201833494},
archivePrefix = {arXiv},
       eprint = {1805.09790},
 primaryClass = {astro-ph.GA},
       adsurl = {https://ui.adsabs.harvard.edu/abs/2018A&A...619A..72R}
}

@ARTICLE{Krumholz2018,
       author = {{Krumholz}, Mark R. and {McKee}, Christopher F. and {Bland-Hawthorn}, Joss},
        title = "{Star Clusters Across Cosmic Time}",
      journal = {\araa},
         year = 2019,
        month = aug,
       volume = {57},
        pages = {227-303},
          doi = {10.1146/annurev-astro-091918-104430},
archivePrefix = {arXiv},
       eprint = {1812.01615},
 primaryClass = {astro-ph.GA},
       adsurl = {https://ui.adsabs.harvard.edu/abs/2019ARA&A..57..227K}
}

@ARTICLE{Hunt2025,
       author = {{Hunt}, Jason A.~S. and {Vasiliev}, Eugene},
        title = "{Milky Way dynamics in light of Gaia}",
      journal = {\nar},
         year = 2025,
        month = jun,
       volume = {100},
          eid = {101721},
        pages = {101721},
          doi = {10.1016/j.newar.2024.101721},
archivePrefix = {arXiv},
       eprint = {2501.04075},
 primaryClass = {astro-ph.GA},
       adsurl = {https://ui.adsabs.harvard.edu/abs/2025NewAR.10001721H}
}

@ARTICLE{Horta2024,
       author = {{Horta}, Danny and {Petersen}, Michael S. and {Pe{\~n}arrubia}, Jorge},
        title = "{Disentangling the Galaxy's Gordian knot: evidence from APOGEE─Gaia for a knotted and slower bar in the Milky Way}",
      journal = {\mnras},
         year = 2025,
        month = apr,
       volume = {538},
       number = {2},
        pages = {998-1018},
          doi = {10.1093/mnras/staf348},
archivePrefix = {arXiv},
       eprint = {2402.07986},
 primaryClass = {astro-ph.GA},
       adsurl = {https://ui.adsabs.harvard.edu/abs/2025MNRAS.538..998H}
}

@ARTICLE{BastianGoodwin2006,
       author = {{Bastian}, N. and {Goodwin}, S.~P.},
        title = "{Evidence for the strong effect of gas removal on the internal dynamics of young stellar clusters}",
      journal = {\mnras},
         year = 2006,
        month = jun,
       volume = {369},
       number = {1},
        pages = {L9-L13},
          doi = {10.1111/j.1745-3933.2006.00162.x},
archivePrefix = {arXiv},
       eprint = {astro-ph/0602465},
 primaryClass = {astro-ph},
       adsurl = {https://ui.adsabs.harvard.edu/abs/2006MNRAS.369L...9B}
}

@ARTICLE{LadaLada2003,
       author = {{Lada}, Charles J. and {Lada}, Elizabeth A.},
        title = "{Embedded Clusters in Molecular Clouds}",
      journal = {\araa},
         year = 2003,
        month = jan,
       volume = {41},
        pages = {57-115},
          doi = {10.1146/annurev.astro.41.011802.094844},
archivePrefix = {arXiv},
       eprint = {astro-ph/0301540},
 primaryClass = {astro-ph},
       adsurl = {https://ui.adsabs.harvard.edu/abs/2003ARA&A..41...57L}
}

@ARTICLE{AllenSantillan,
       author = {{Allen}, Christine and {Santillan}, Alfredo},
        title = "{An improved model of the galactic mass distribution for orbit computations.}",
      journal = {\rmxaa},
         year = 1991,
        month = oct,
       volume = {22},
        pages = {255},
       adsurl = {https://ui.adsabs.harvard.edu/abs/1991RMxAA..22..255A}
}

@ARTICLE{LongMurali92,
       author = {{Long}, Kevin and {Murali}, Chigurupati},
        title = "{Analytical Potentials for Barred Galaxies}",
      journal = {\apj},
         year = 1992,
        month = sep,
       volume = {397},
        pages = {44},
          doi = {10.1086/171764},
       adsurl = {https://ui.adsabs.harvard.edu/abs/1992ApJ...397...44L}
}

@BOOK{BinneyTremaine2008,
       author = {{Binney}, James and {Tremaine}, Scott},
        title = "{Galactic Dynamics: Second Edition}",
         year = 2008,
       adsurl = {https://ui.adsabs.harvard.edu/abs/2008gady.book.....B}
}

@ARTICLE{Reid2014,
       author = {{Reid}, M.~J. and {Menten}, K.~M. and {Brunthaler}, A. and {Zheng}, X.~W. and {Dame}, T.~M. and {Xu}, Y. and {Wu}, Y. and {Zhang}, B. and {Sanna}, A. and {Sato}, M. and {Hachisuka}, K. and {Choi}, Y.~K. and {Immer}, K. and {Moscadelli}, L. and {Rygl}, K.~L.~J. and {Bartkiewicz}, A.},
        title = "{Trigonometric Parallaxes of High Mass Star Forming Regions: The Structure and Kinematics of the Milky Way}",
      journal = {\apj},
         year = 2014,
        month = mar,
       volume = {783},
       number = {2},
          eid = {130},
        pages = {130},
          doi = {10.1088/0004-637X/783/2/130},
archivePrefix = {arXiv},
       eprint = {1401.5377},
 primaryClass = {astro-ph.GA},
       adsurl = {https://ui.adsabs.harvard.edu/abs/2014ApJ...783..130R}
}

@ARTICLE{Chen2001,
       author = {{Chen}, Bing and {Stoughton}, Chris and {Smith}, J. Allyn and {Uomoto}, Alan and {Pier}, Jeffrey R. and {Yanny}, Brian and {Ivezi{\'c}}, {\v{Z}}eljko and {York}, Donald G. and {Anderson}, John E. and {Annis}, James and {Brinkmann}, Jon and {Csabai}, Istv{\'a}n and {Fukugita}, Masataka and {Hindsley}, Robert and {Lupton}, Robert and {Munn}, Jeffrey A. and {SDSS Collaboration}},
        title = "{Stellar Population Studies with the SDSS. I. The Vertical Distribution of Stars in the Milky Way}",
      journal = {\apj},
         year = 2001,
        month = may,
       volume = {553},
       number = {1},
        pages = {184-197},
          doi = {10.1086/320647},
       adsurl = {https://ui.adsabs.harvard.edu/abs/2001ApJ...553..184C}
}

@ARTICLE{Schonrich2010,
       author = {{Sch{\"o}nrich}, Ralph and {Binney}, James and {Dehnen}, Walter},
        title = "{Local kinematics and the local standard of rest}",
      journal = {\mnras},
         year = 2010,
        month = apr,
       volume = {403},
       number = {4},
        pages = {1829-1833},
          doi = {10.1111/j.1365-2966.2010.16253.x},
archivePrefix = {arXiv},
       eprint = {0912.3693},
 primaryClass = {astro-ph.GA},
       adsurl = {https://ui.adsabs.harvard.edu/abs/2010MNRAS.403.1829S}
}

@ARTICLE{galpy,
       author = {{Bovy}, Jo},
        title = "{galpy: A python Library for Galactic Dynamics}",
      journal = {\apjs},
         year = 2015,
        month = feb,
       volume = {216},
       number = {2},
          eid = {29},
        pages = {29},
          doi = {10.1088/0067-0049/216/2/29},
archivePrefix = {arXiv},
       eprint = {1412.3451},
 primaryClass = {astro-ph.GA},
       adsurl = {https://ui.adsabs.harvard.edu/abs/2015ApJS..216...29B}
}

@ARTICLE{Kupper2010,
       author = {{K{\"u}pper}, Andreas H.~W. and {Kroupa}, Pavel and {Baumgardt}, Holger and {Heggie}, Douglas C.},
        title = "{Tidal tails of star clusters}",
      journal = {\mnras},
         year = 2010,
        month = jan,
       volume = {401},
       number = {1},
        pages = {105-120},
          doi = {10.1111/j.1365-2966.2009.15690.x},
archivePrefix = {arXiv},
       eprint = {0909.2619},
 primaryClass = {astro-ph.SR},
       adsurl = {https://ui.adsabs.harvard.edu/abs/2010MNRAS.401..105K}
}

@ARTICLE{Just2008,
       author = {{Just}, A. and {Berczik}, P. and {Petrov}, M.~I. and {Ernst}, A.},
        title = "{Quantitative analysis of clumps in the tidal tails of star clusters}",
      journal = {\mnras},
         year = 2009,
        month = jan,
       volume = {392},
       number = {3},
        pages = {969-981},
          doi = {10.1111/j.1365-2966.2008.14099.x},
archivePrefix = {arXiv},
       eprint = {0808.3293},
 primaryClass = {astro-ph},
       adsurl = {https://ui.adsabs.harvard.edu/abs/2009MNRAS.392..969J}
}

@ARTICLE{Kupper2008,
       author = {{K{\"u}pper}, Andreas H.~W. and {MacLeod}, Andrew and {Heggie}, Douglas C.},
        title = "{On the structure of tidal tails}",
      journal = {\mnras},
         year = 2008,
        month = jul,
       volume = {387},
       number = {3},
        pages = {1248-1252},
          doi = {10.1111/j.1365-2966.2008.13323.x},
archivePrefix = {arXiv},
       eprint = {0804.2476},
 primaryClass = {astro-ph},
       adsurl = {https://ui.adsabs.harvard.edu/abs/2008MNRAS.387.1248K}
}

@ARTICLE{Kupper2012,
       author = {{K{\"u}pper}, Andreas H.~W. and {Lane}, Richard R. and {Heggie}, Douglas C.},
        title = "{More on the structure of tidal tails}",
      journal = {\mnras},
         year = 2012,
        month = mar,
       volume = {420},
       number = {3},
        pages = {2700-2714},
          doi = {10.1111/j.1365-2966.2011.20242.x},
archivePrefix = {arXiv},
       eprint = {1111.5013},
 primaryClass = {astro-ph.GA},
       adsurl = {https://ui.adsabs.harvard.edu/abs/2012MNRAS.420.2700K}
}

@ARTICLE{ContopoulosGrosbol1989,
       author = {{Contopoulos}, G. and {Grosbol}, P.},
        title = "{Orbits in barred galaxies}",
      journal = {\aapr},
         year = 1989,
        month = nov,
       volume = {1},
       number = {3-4},
        pages = {261-289},
          doi = {10.1007/BF00873080},
       adsurl = {https://ui.adsabs.harvard.edu/abs/1989A&ARv...1..261C}
}

@ARTICLE{elpigraph,
       author = {{Albergante}, Luca and {Mirkes}, Evgeny and {Bac}, Jonathan and {Chen}, Huidong and {Martin}, Alexis and {Faure}, Louis and {Barillot}, Emmanuel and {Pinello}, Luca and {Gorban}, Alexander and {Zinovyev}, Andrei},
        title = "{Robust and Scalable Learning of Complex Intrinsic Dataset Geometry via ElPiGraph}",
      journal = {Entropy},
         year = 2020,
        month = mar,
       volume = {22},
       number = {3},
          eid = {296},
        pages = {296},
          doi = {10.3390/e22030296},
archivePrefix = {arXiv},
       eprint = {1804.07580},
 primaryClass = {cs.LG},
       adsurl = {https://ui.adsabs.harvard.edu/abs/2020Entrp..22..296A}
}

@ARTICLE{Tarricq22,
       author = {{Tarricq}, Y. and {Soubiran}, C. and {Casamiquela}, L. and {Castro-Ginard}, A. and {Olivares}, J. and {Miret-Roig}, N. and {Galli}, P.~A.~B.},
        title = "{Structural parameters of 389 local open clusters}",
      journal = {\aap},
         year = 2022,
        month = mar,
       volume = {659},
          eid = {A59},
        pages = {A59},
          doi = {10.1051/0004-6361/202142186},
archivePrefix = {arXiv},
       eprint = {2111.05291},
 primaryClass = {astro-ph.GA},
       adsurl = {https://ui.adsabs.harvard.edu/abs/2022A&A...659A..59T}
}

@ARTICLE{Zhou2026,
       author = {{Zhou}, Zi-yi and {Wang}, Long and {Jerabkova}, Tereza and {He}, Zhenghao},
        title = "{Constraining the Galactic Bar and Spiral Pattern Speeds with the Hyades Tidal Stream}",
      journal = {\apj},
         year = 2026,
        month = may,
       volume = {1002},
       number = {1},
          eid = {37},
        pages = {37},
          doi = {10.3847/1538-4357/ae5959},
       adsurl = {https://ui.adsabs.harvard.edu/abs/2026ApJ..1002...37Z}
}

@ARTICLE{Tang2019,
       author = {{Tang}, Shih-Yun and {Pang}, Xiaoying and {Yuan}, Zhen and {Chen}, W.~P. and {Hong}, Jongsuk and {Goldman}, Bertrand and {Just}, Andreas and {Shukirgaliyev}, Bekdaulet and {Lin}, Chien-Cheng},
        title = "{Discovery of Tidal Tails in Disrupting Open Clusters: Coma Berenices and a Neighbor Stellar Group}",
      journal = {\apj},
         year = 2019,
        month = may,
       volume = {877},
       number = {1},
          eid = {12},
        pages = {12},
          doi = {10.3847/1538-4357/ab13b0},
archivePrefix = {arXiv},
       eprint = {1902.01404},
 primaryClass = {astro-ph.GA},
       adsurl = {https://ui.adsabs.harvard.edu/abs/2019ApJ...877...12T}
}

@ARTICLE{Luccini2024,
       author = {{Lucchini}, Scott and {D'Onghia}, Elena and {Aguerri}, J. Alfonso L.},
        title = "{The Milky Way bar pattern speed using Hercules and Gaia DR3}",
      journal = {\mnras},
         year = 2024,
        month = jun,
       volume = {531},
       number = {1},
        pages = {L14-L19},
          doi = {10.1093/mnrasl/slae024},
archivePrefix = {arXiv},
       eprint = {2305.04981},
 primaryClass = {astro-ph.GA},
       adsurl = {https://ui.adsabs.harvard.edu/abs/2024MNRAS.531L..14L}
}

@ARTICLE{Bovy2019,
       author = {{Bovy}, Jo and {Leung}, Henry W. and {Hunt}, Jason A.~S. and {Mackereth}, J. Ted and {Garc{\'\i}a-Hern{\'a}ndez}, Domingo A. and {Roman-Lopes}, Alexandre},
        title = "{Life in the fast lane: a direct view of the dynamics, formation, and evolution of the Milky Way's bar}",
      journal = {\mnras},
         year = 2019,
        month = dec,
       volume = {490},
       number = {4},
        pages = {4740-4747},
          doi = {10.1093/mnras/stz2891},
archivePrefix = {arXiv},
       eprint = {1905.11404},
 primaryClass = {astro-ph.GA},
       adsurl = {https://ui.adsabs.harvard.edu/abs/2019MNRAS.490.4740B}
}

@ARTICLE{Hunt2018b,
       author = {{Hunt}, Jason A.~S. and {Bovy}, Jo},
        title = "{The 4:1 outer Lindblad resonance of a long-slow bar as an explanation for the Hercules stream}",
      journal = {\mnras},
         year = 2018,
        month = jul,
       volume = {477},
       number = {3},
        pages = {3945-3953},
          doi = {10.1093/mnras/sty921},
archivePrefix = {arXiv},
       eprint = {1803.02358},
 primaryClass = {astro-ph.GA},
       adsurl = {https://ui.adsabs.harvard.edu/abs/2018MNRAS.477.3945H}
}

@ARTICLE{Kos2026,
       author = {{Kos}, Janez and {Risojevi{\'c}}, Jovana and {Ilc}, Samo},
        title = "{Dynamics of tidal tails of open clusters: I. effects of bar, spiral arms and giant molecular clouds}",
      journal = {arXiv e-prints},
         year = 2026,
        month = may,
          eid = {arXiv:2605.31439},
        pages = {arXiv:2605.31439},
archivePrefix = {arXiv},
       eprint = {2605.31439},
 primaryClass = {astro-ph.GA},
       adsurl = {https://ui.adsabs.harvard.edu/abs/2026arXiv260531439K}
}

@ARTICLE{Zhang2020,
       author = {{Zhang}, Yu and {Tang}, Shih-Yun and {Chen}, W.~P. and {Pang}, Xiaoying and {Liu}, J.~Z.},
        title = "{Diagnosing the Stellar Population and Tidal Structure of the Blanco 1 Star Cluster}",
      journal = {\apj},
         year = 2020,
        month = feb,
       volume = {889},
       number = {2},
          eid = {99},
        pages = {99},
          doi = {10.3847/1538-4357/ab63d4},
archivePrefix = {arXiv},
       eprint = {1912.06657},
 primaryClass = {astro-ph.GA},
       adsurl = {https://ui.adsabs.harvard.edu/abs/2020ApJ...889...99Z}
}

@ARTICLE{Pang2022,
       author = {{Pang}, Xiaoying and {Tang}, Shih-Yun and {Li}, Yuqian and {Yu}, Zeqiu and {Wang}, Long and {Li}, Jiayu and {Li}, Yezhang and {Wang}, Yifan and {Wang}, Yanshu and {Zhang}, Teng and {Pasquato}, Mario and {Kouwenhoven}, M.~B.~N.},
        title = "{3D Morphology of Open Clusters in the Solar Neighborhood with Gaia EDR 3. II. Hierarchical Star Formation Revealed by Spatial and Kinematic Substructures}",
      journal = {\apj},
         year = 2022,
        month = jun,
       volume = {931},
       number = {2},
          eid = {156},
        pages = {156},
          doi = {10.3847/1538-4357/ac674e},
archivePrefix = {arXiv},
       eprint = {2204.06000},
 primaryClass = {astro-ph.GA},
       adsurl = {https://ui.adsabs.harvard.edu/abs/2022ApJ...931..156P}
}

@ARTICLE{Jadhav25,
       author = {{Jadhav}, Vikrant V. and {Risbud}, Dhanraj and {Kroupa}, Pavel and {Wu}, Wenjie},
        title = "{Tidal tails of nearby open clusters: II. A review of simulated properties and the reliability of observational catalogues}",
      journal = {\aap},
         year = 2025,
        month = dec,
       volume = {704},
          eid = {A50},
        pages = {A50},
          doi = {10.1051/0004-6361/202555858},
archivePrefix = {arXiv},
       eprint = {2508.15056},
 primaryClass = {astro-ph.GA},
       adsurl = {https://ui.adsabs.harvard.edu/abs/2025A&A...704A..50J}
}

@ARTICLE{Kos24,
       author = {{Kos}, Janez},
        title = "{Tidal tails of open clusters}",
      journal = {\aap},
         year = 2024,
        month = nov,
       volume = {691},
          eid = {A28},
        pages = {A28},
          doi = {10.1051/0004-6361/202449828},
archivePrefix = {arXiv},
       eprint = {2406.18767},
 primaryClass = {astro-ph.GA},
       adsurl = {https://ui.adsabs.harvard.edu/abs/2024A&A...691A..28K}
}

@ARTICLE{Vaher23,
       author = {{Vaher}, E. and {Hobbs}, D. and {McMillan}, P. and {Prusti}, T.},
        title = "{Finding the dispersing siblings of young open clusters. Dynamical traceback simulations using Gaia DR3}",
      journal = {\aap},
         year = 2023,
        month = nov,
       volume = {679},
          eid = {A105},
        pages = {A105},
          doi = {10.1051/0004-6361/202346877},
archivePrefix = {arXiv},
       eprint = {2310.02441},
 primaryClass = {astro-ph.GA},
       adsurl = {https://ui.adsabs.harvard.edu/abs/2023A&A...679A.105V}
}

@ARTICLE{Risbud25,
       author = {{Risbud}, Dhanraj and {Jadhav}, Vikrant V. and {Kroupa}, Pavel},
        title = "{Tidal tails of nearby open clusters: I. Mapping with Gaia DR3}",
      journal = {\aap},
         year = 2025,
        month = feb,
       volume = {694},
          eid = {A258},
        pages = {A258},
          doi = {10.1051/0004-6361/202453302},
archivePrefix = {arXiv},
       eprint = {2501.17225},
 primaryClass = {astro-ph.GA},
       adsurl = {https://ui.adsabs.harvard.edu/abs/2025A&A...694A.258R}
}

@ARTICLE{Ferrone2025,
       author = {{Ferrone}, Salvatore and {Montuori}, Marco and {Di Matteo}, Paola and {Mastrobuono-Battisti}, Alessandra and {Ibata}, Rodrigo and {Bianchini}, Paolo and {Khoperskov}, Sergey and {Leclerc}, Nicolas and {Hottier}, Clement and {Stein}, Eliot and {Valls-Gabaud}, David and {Owain Snaith}, N. and {Haywood}, Misha},
        title = "{Gaps in stellar streams as a result of globular cluster flybys: The case of Palomar 5}",
      journal = {\aap},
         year = 2025,
        month = jul,
       volume = {699},
          eid = {A289},
        pages = {A289},
          doi = {10.1051/0004-6361/202553923},
archivePrefix = {arXiv},
       eprint = {2502.03941},
 primaryClass = {astro-ph.GA},
       adsurl = {https://ui.adsabs.harvard.edu/abs/2025A&A...699A.289F}
}

@ARTICLE{Ferrone2023,
       author = {{Ferrone}, Salvatore and {Di Matteo}, Paola and {Mastrobuono-Battisti}, Alessandra and {Haywood}, Misha and {Snaith}, Owain N. and {Montuori}, Marco and {Khoperskov}, Sergey and {Valls-Gabaud}, David},
        title = "{The e-TidalGCs project. Modeling the extra-tidal features generated by Galactic globular clusters}",
      journal = {\aap},
         year = 2023,
        month = may,
       volume = {673},
          eid = {A44},
        pages = {A44},
          doi = {10.1051/0004-6361/202244141},
archivePrefix = {arXiv},
       eprint = {2301.05166},
 primaryClass = {astro-ph.GA},
       adsurl = {https://ui.adsabs.harvard.edu/abs/2023A&A...673A..44F}
}

@ARTICLE{Thomas2023,
       author = {{Thomas}, Guillaume F. and {Famaey}, Benoit and {Monari}, Giacomo and {Laporte}, Chervin F.~P. and {Ibata}, Rodrigo and {de Laverny}, Patrick and {Hill}, Vanessa and {Boily}, Christian},
        title = "{Impact of the Galactic bar on tidal streams within the Galactic disc. The case of the tidal stream of the Hyades}",
      journal = {\aap},
         year = 2023,
        month = oct,
       volume = {678},
          eid = {A180},
        pages = {A180},
          doi = {10.1051/0004-6361/202346650},
archivePrefix = {arXiv},
       eprint = {2309.05733},
 primaryClass = {astro-ph.GA},
       adsurl = {https://ui.adsabs.harvard.edu/abs/2023A&A...678A.180T}
}

@ARTICLE{Bhattacharya22,
       author = {{Bhattacharya}, Souradeep and {Rao}, Khushboo K. and {Agarwal}, Manan and {Balan}, Shanmugha and {Vaidya}, Kaushar},
        title = "{A Gaia EDR3 search for tidal tails in disintegrating open clusters}",
      journal = {\mnras},
         year = 2022,
        month = dec,
       volume = {517},
       number = {3},
        pages = {3525-3549},
          doi = {10.1093/mnras/stac2906},
archivePrefix = {arXiv},
       eprint = {2209.08259},
 primaryClass = {astro-ph.GA},
       adsurl = {https://ui.adsabs.harvard.edu/abs/2022MNRAS.517.3525B}
}

@ARTICLE{Boffin22,
       author = {{Boffin}, Henri M.~J. and {Jerabkova}, Tereza and {Beccari}, Giacomo and {Wang}, Long},
        title = "{A tale of caution: the tails of NGC 752 are much longer than claimed}",
      journal = {\mnras},
         year = 2022,
        month = aug,
       volume = {514},
       number = {3},
        pages = {3579-3592},
          doi = {10.1093/mnras/stac1567},
archivePrefix = {arXiv},
       eprint = {2205.11949},
 primaryClass = {astro-ph.GA},
       adsurl = {https://ui.adsabs.harvard.edu/abs/2022MNRAS.514.3579B}
}

@ARTICLE{Jerabkova21,
       author = {{Jerabkova}, Tereza and {Boffin}, Henri M.~J. and {Beccari}, Giacomo and {de Marchi}, Guido and {de Bruijne}, Jos H.~J. and {Prusti}, Timo},
        title = "{The 800 pc long tidal tails of the Hyades star cluster. Possible discovery of candidate epicyclic overdensities from an open star cluster}",
      journal = {\aap},
         year = 2021,
        month = mar,
       volume = {647},
          eid = {A137},
        pages = {A137},
          doi = {10.1051/0004-6361/202039949},
archivePrefix = {arXiv},
       eprint = {2103.12080},
 primaryClass = {astro-ph.GA},
       adsurl = {https://ui.adsabs.harvard.edu/abs/2021A&A...647A.137J}
}

@ARTICLE{Miller2025,
       author = {{Miller}, Alexis N. and {Tregoning}, Kyle R. and {Andrews}, Jeff J. and {Schuler}, Simon C. and {Curtis}, Jason L. and {Ag{\"u}eros}, Marcel A. and {Cargile}, Phillip A. and {Chanam{\'e}}, Julio},
        title = "{Evidence for a Catastrophically Disrupted Open Cluster}",
      journal = {\apj},
         year = 2025,
        month = jun,
       volume = {986},
       number = {1},
          eid = {27},
        pages = {27},
          doi = {10.3847/1538-4357/adceb8},
archivePrefix = {arXiv},
       eprint = {2504.19343},
 primaryClass = {astro-ph.GA},
       adsurl = {https://ui.adsabs.harvard.edu/abs/2025ApJ...986...27M}
}

@ARTICLE{Irrgang2013,
       author = {{Irrgang}, A. and {Wilcox}, B. and {Tucker}, E. and {Schiefelbein}, L.},
        title = "{Milky Way mass models for orbit calculations}",
      journal = {\aap},
         year = 2013,
        month = jan,
       volume = {549},
          eid = {A137},
        pages = {A137},
          doi = {10.1051/0004-6361/201220540},
archivePrefix = {arXiv},
       eprint = {1211.4353},
 primaryClass = {astro-ph.GA},
       adsurl = {https://ui.adsabs.harvard.edu/abs/2013A&A...549A.137I}
}

@ARTICLE{Wiggins25,
       author = {{Wiggins}, Alessa I. and {Quinn}, Jamie R. and {Oeur}, Micah and {Loebman}, Sarah R. and {Frinchaboy}, Peter M. and {Daniel}, Kathryne J. and {McCluskey}, Fiona and {Otto}, Jonah M. and {Woodward}, Hannah R. and {D'Onghia}, Elena and {Wetzel}, Andrew and {Parul}, Hanna and {Bhattarai}, Binod and {Cozzi}, Maximilian},
        title = "{Understanding the Origin and Dynamical Evolution of the Unique Open Star Cluster Berkeley 20 Using FIRE Simulations}",
      journal = {\apjl},
         year = 2025,
        month = dec,
       volume = {995},
       number = {1},
          eid = {L25},
        pages = {L25},
          doi = {10.3847/2041-8213/ae21bf},
archivePrefix = {arXiv},
       eprint = {2511.14958},
 primaryClass = {astro-ph.GA},
       adsurl = {https://ui.adsabs.harvard.edu/abs/2025ApJ...995L..25W}
}

@ARTICLE{Meingast19,
       author = {{Meingast}, Stefan and {Alves}, Jo{\~a}o},
        title = "{Extended stellar systems in the solar neighborhood. I. The tidal tails of the Hyades}",
      journal = {\aap},
         year = 2019,
        month = jan,
       volume = {621},
          eid = {L3},
        pages = {L3},
          doi = {10.1051/0004-6361/201834622},
archivePrefix = {arXiv},
       eprint = {1811.04931},
 primaryClass = {astro-ph.GA},
       adsurl = {https://ui.adsabs.harvard.edu/abs/2019A&A...621L...3M}
}

@ARTICLE{Meingast21,
       author = {{Meingast}, Stefan and {Alves}, Jo{\~a}o and {Rottensteiner}, Alena},
        title = "{Extended stellar systems in the solar neighborhood. V. Discovery of coronae of nearby star clusters}",
      journal = {\aap},
         year = 2021,
        month = jan,
       volume = {645},
          eid = {A84},
        pages = {A84},
          doi = {10.1051/0004-6361/202038610},
archivePrefix = {arXiv},
       eprint = {2010.06591},
 primaryClass = {astro-ph.GA},
       adsurl = {https://ui.adsabs.harvard.edu/abs/2021A&A...645A..84M}
}

@ARTICLE{Chumak2006,
       author = {{Chumak}, Ya. O. and {Rastorguev}, A.~S.},
        title = "{Analysis of the structure and dynamics of the stellar tails of open star clusters}",
      journal = {Astronomy Letters},
         year = 2006,
        month = mar,
       volume = {32},
       number = {3},
        pages = {157-165},
          doi = {10.1134/S1063773706030030},
       adsurl = {https://ui.adsabs.harvard.edu/abs/2006AstL...32..157C}
}

@ARTICLE{Dinnbier20,
       author = {{Dinnbier}, Franti{\v{s}}ek and {Kroupa}, Pavel},
        title = "{Tidal tails of open star clusters as probes of early gas expulsion. I. A semi-analytic model}",
      journal = {\aap},
         year = 2020,
        month = aug,
       volume = {640},
          eid = {A84},
        pages = {A84},
          doi = {10.1051/0004-6361/201936570},
archivePrefix = {arXiv},
       eprint = {2006.14087},
 primaryClass = {astro-ph.GA},
       adsurl = {https://ui.adsabs.harvard.edu/abs/2020A&A...640A..84D}
}

@ARTICLE{Roser19b,
       author = {{R{\"o}ser}, Siegfried and {Schilbach}, Elena},
        title = "{Praesepe (NGC 2632) and its tidal tails}",
      journal = {\aap},
         year = 2019,
        month = jul,
       volume = {627},
          eid = {A4},
        pages = {A4},
          doi = {10.1051/0004-6361/201935502},
archivePrefix = {arXiv},
       eprint = {1903.08610},
 primaryClass = {astro-ph.SR},
       adsurl = {https://ui.adsabs.harvard.edu/abs/2019A&A...627A...4R}
}

@ARTICLE{Roser19a,
       author = {{R{\"o}ser}, Siegfried and {Schilbach}, Elena and {Goldman}, Bertrand},
        title = "{Hyades tidal tails revealed by Gaia DR2}",
      journal = {\aap},
         year = 2019,
        month = jan,
       volume = {621},
          eid = {L2},
        pages = {L2},
          doi = {10.1051/0004-6361/201834608},
archivePrefix = {arXiv},
       eprint = {1811.03845},
 primaryClass = {astro-ph.SR},
       adsurl = {https://ui.adsabs.harvard.edu/abs/2019A&A...621L...2R}
}

@ARTICLE{2024MNRAS.528.5189G,
       author = {{Grondin}, Steffani M. and {Webb}, Jeremy J. and {Lane}, James M.~M. and {Speagle}, Joshua S. and {Leigh}, Nathan W.~C.},
        title = "{A catalogue of Galactic GEMS: Globular cluster Extra-tidal Mock Stars}",
      journal = {\mnras},
         year = 2024,
        month = mar,
       volume = {528},
       number = {3},
        pages = {5189-5211},
          doi = {10.1093/mnras/stae203},
archivePrefix = {arXiv},
       eprint = {2310.09331},
 primaryClass = {astro-ph.GA},
       adsurl = {https://ui.adsabs.harvard.edu/abs/2024MNRAS.528.5189G}
}

@ARTICLE{2024ApJ...967...89I,
       author = {{Ibata}, Rodrigo and {Malhan}, Khyati and {Tenachi}, Wassim and {Ardern-Arentsen}, Anke and {Bellazzini}, Michele and {Bianchini}, Paolo and {Bonifacio}, Piercarlo and {Caffau}, Elisabetta and {Diakogiannis}, Foivos and {Errani}, Raphael and {Famaey}, Benoit and {Ferrone}, Salvatore and {Martin}, Nicolas F. and {di Matteo}, Paola and {Monari}, Giacomo and {Renaud}, Florent and {Starkenburg}, Else and {Thomas}, Guillaume and {Viswanathan}, Akshara and {Yuan}, Zhen},
        title = "{Charting the Galactic Acceleration Field. II. A Global Mass Model of the Milky Way from the STREAMFINDER Atlas of Stellar Streams Detected in Gaia DR3}",
      journal = {\apj},
         year = 2024,
        month = jun,
       volume = {967},
       number = {2},
          eid = {89},
        pages = {89},
          doi = {10.3847/1538-4357/ad382d},
archivePrefix = {arXiv},
       eprint = {2311.17202},
 primaryClass = {astro-ph.GA},
       adsurl = {https://ui.adsabs.harvard.edu/abs/2024ApJ...967...89I}
}

@ARTICLE{2007ApJ...659.1212M,
       author = {{Montuori}, M. and {Capuzzo-Dolcetta}, R. and {Di Matteo}, P. and {Lepinette}, A. and {Miocchi}, P.},
        title = "{Tidal Tails around Globular Clusters: Are They a Good Tracer of Cluster Orbits?}",
      journal = {\apj},
         year = 2007,
        month = apr,
       volume = {659},
       number = {2},
        pages = {1212-1221},
          doi = {10.1086/512114},
archivePrefix = {arXiv},
       eprint = {astro-ph/0611204},
 primaryClass = {astro-ph},
       adsurl = {https://ui.adsabs.harvard.edu/abs/2007ApJ...659.1212M}
}

@ARTICLE{2012A&A...546L...7M,
       author = {{Mastrobuono-Battisti}, A. and {Di Matteo}, P. and {Montuori}, M. and {Haywood}, M.},
        title = "{Clumpy streams in a smooth dark halo: the case of Palomar 5}",
      journal = {\aap},
         year = 2012,
        month = oct,
       volume = {546},
          eid = {L7},
        pages = {L7},
          doi = {10.1051/0004-6361/201219563},
archivePrefix = {arXiv},
       eprint = {1209.0466},
 primaryClass = {astro-ph.GA},
       adsurl = {https://ui.adsabs.harvard.edu/abs/2012A&A...546L...7M}
}

@ARTICLE{2025NewAR.10001713B,
       author = {{Bonaca}, Ana and {Price-Whelan}, Adrian M.},
        title = "{Stellar streams in the Gaia era}",
      journal = {\nar},
         year = 2025,
        month = jun,
       volume = {100},
          eid = {101713},
        pages = {101713},
          doi = {10.1016/j.newar.2024.101713},
archivePrefix = {arXiv},
       eprint = {2405.19410},
 primaryClass = {astro-ph.GA},
       adsurl = {https://ui.adsabs.harvard.edu/abs/2025NewAR.10001713B}
}

@ARTICLE{2004AJ....127.2753D,
       author = {{Dehnen}, Walter and {Odenkirchen}, Michael and {Grebel}, Eva K. and {Rix}, Hans-Walter},
        title = "{Modeling the Disruption of the Globular Cluster Palomar 5 by Galactic Tides}",
      journal = {\aj},
         year = 2004,
        month = may,
       volume = {127},
       number = {5},
        pages = {2753-2770},
          doi = {10.1086/383214},
archivePrefix = {arXiv},
       eprint = {astro-ph/0401422},
 primaryClass = {astro-ph},
       adsurl = {https://ui.adsabs.harvard.edu/abs/2004AJ....127.2753D}
}

@ARTICLE{2024NewAR..9901696C,
       author = {{Cantat-Gaudin}, T. and {Casamiquela}, L.},
        title = "{How Gaia sheds light on the Milky Way star cluster population}",
      journal = {\nar},
         year = 2024,
        month = dec,
       volume = {99},
          eid = {101696},
        pages = {101696},
          doi = {10.1016/j.newar.2024.101696},
archivePrefix = {arXiv},
       eprint = {2406.03308},
 primaryClass = {astro-ph.GA},
       adsurl = {https://ui.adsabs.harvard.edu/abs/2024NewAR..9901696C}
}

@article{gnedin_destruction_1997,
	title = {Destruction of the {Galactic} {Globular} {Cluster} {System}},
	volume = {474},
	issn = {0004-637X},
	url = {https://iopscience.iop.org/article/10.1086/303441},
	doi = {10.1086/303441},
	language = {en},
	number = {1},
	urldate = {2026-06-02},
	journal = {The Astrophysical Journal},
	publisher = {IOP Publishing},
	author = {Gnedin, Oleg Y. and Ostriker, Jeremiah P.},
	month = jan,
	year = {1997},
	pages = {223},
}

@ARTICLE{2017MNRAS.465...76M,
       author = {{McMillan}, Paul J.},
        title = "{The mass distribution and gravitational potential of the Milky Way}",
      journal = {\mnras},
         year = 2017,
        month = feb,
       volume = {465},
       number = {1},
        pages = {76-94},
          doi = {10.1093/mnras/stw2759},
archivePrefix = {arXiv},
       eprint = {1608.00971},
 primaryClass = {astro-ph.GA},
       adsurl = {https://ui.adsabs.harvard.edu/abs/2017MNRAS.465...76M}
}

@ARTICLE{1998MNRAS.294..429D,
       author = {{Dehnen}, Walter and {Binney}, James},
        title = "{Mass models of the Milky Way}",
      journal = {\mnras},
         year = 1998,
        month = mar,
       volume = {294},
       number = {3},
        pages = {429-438},
          doi = {10.1046/j.1365-8711.1998.01282.x10.1111/j.1365-8711.1998.01282.x},
archivePrefix = {arXiv},
       eprint = {astro-ph/9612059},
 primaryClass = {astro-ph},
       adsurl = {https://ui.adsabs.harvard.edu/abs/1998MNRAS.294..429D}
}

@ARTICLE{2013A&A...549A.137I,
       author = {{Irrgang}, A. and {Wilcox}, B. and {Tucker}, E. and {Schiefelbein}, L.},
        title = "{Milky Way mass models for orbit calculations}",
      journal = {\aap},
         year = 2013,
        month = jan,
       volume = {549},
          eid = {A137},
        pages = {A137},
          doi = {10.1051/0004-6361/201220540},
archivePrefix = {arXiv},
       eprint = {1211.4353},
 primaryClass = {astro-ph.GA},
       adsurl = {https://ui.adsabs.harvard.edu/abs/2013A&A...549A.137I}
}

@ARTICLE{1991RMxAA..22..255A,
       author = {{Allen}, Christine and {Santillan}, Alfredo},
        title = "{An improved model of the galactic mass distribution for orbit computations.}",
      journal = {\rmxaa},
         year = 1991,
        month = oct,
       volume = {22},
        pages = {255},
       adsurl = {https://ui.adsabs.harvard.edu/abs/1991RMxAA..22..255A}}

@ARTICLE{2010ApJ...712..260K,
       author = {{Koposov}, Sergey E. and {Rix}, Hans-Walter and {Hogg}, David W.},
        title = "{Constraining the Milky Way Potential with a Six-Dimensional Phase-Space Map of the GD-1 Stellar Stream}",
      journal = {\apj},
         year = 2010,
        month = mar,
       volume = {712},
       number = {1},
        pages = {260-273},
          doi = {10.1088/0004-637X/712/1/260},
archivePrefix = {arXiv},
       eprint = {0907.1085},
 primaryClass = {astro-ph.GA},
       adsurl = {https://ui.adsabs.harvard.edu/abs/2010ApJ...712..260K}
}

@ARTICLE{2011MNRAS.417..198V,
       author = {{Varghese}, A. and {Ibata}, R. and {Lewis}, G.~F.},
        title = "{Stellar streams as probes of dark halo mass and morphology: a Bayesian reconstruction}",
      journal = {\mnras},
         year = 2011,
        month = oct,
       volume = {417},
       number = {1},
        pages = {198-215},
          doi = {10.1111/j.1365-2966.2011.19097.x},
archivePrefix = {arXiv},
       eprint = {1106.1765},
 primaryClass = {astro-ph.GA},
       adsurl = {https://ui.adsabs.harvard.edu/abs/2011MNRAS.417..198V}
}

@ARTICLE{2016ApJ...833...31B,
       author = {{Bovy}, Jo and {Bahmanyar}, Anita and {Fritz}, Tobias K. and {Kallivayalil}, Nitya},
        title = "{The Shape of the Inner Milky Way Halo from Observations of the Pal 5 and GD--1 Stellar Streams}",
      journal = {\apj},
         year = 2016,
        month = dec,
       volume = {833},
       number = {1},
          eid = {31},
        pages = {31},
          doi = {10.3847/1538-4357/833/1/31},
archivePrefix = {arXiv},
       eprint = {1609.01298},
 primaryClass = {astro-ph.GA},
       adsurl = {https://ui.adsabs.harvard.edu/abs/2016ApJ...833...31B}
}

@ARTICLE{2010ApJ...714..229L,
       author = {{Law}, David R. and {Majewski}, Steven R.},
        title = "{The Sagittarius Dwarf Galaxy: A Model for Evolution in a Triaxial Milky Way Halo}",
      journal = {\apj},
         year = 2010,
        month = may,
       volume = {714},
       number = {1},
        pages = {229-254},
          doi = {10.1088/0004-637X/714/1/229},
archivePrefix = {arXiv},
       eprint = {1003.1132},
 primaryClass = {astro-ph.GA},
       adsurl = {https://ui.adsabs.harvard.edu/abs/2010ApJ...714..229L}
}

@ARTICLE{2023ApJ...953...19C,
       author = {{Cabrera}, Tom{\'a}s and {Rodriguez}, Carl L.},
        title = "{Runaway and Hypervelocity Stars from Compact Object Encounters in Globular Clusters}",
      journal = {\apj},
         year = 2023,
        month = aug,
       volume = {953},
       number = {1},
          eid = {19},
        pages = {19},
          doi = {10.3847/1538-4357/acdc22},
archivePrefix = {arXiv},
       eprint = {2302.03048},
 primaryClass = {astro-ph.GA},
       adsurl = {https://ui.adsabs.harvard.edu/abs/2023ApJ...953...19C}
}

@ARTICLE{2005AJ....129.1906C,
       author = {{Capuzzo Dolcetta}, R. and {Di Matteo}, P. and {Miocchi}, P.},
        title = "{Formation and Evolution of Clumpy Tidal Tails around Globular Clusters}",
      journal = {\aj},
         year = 2005,
        month = apr,
       volume = {129},
       number = {4},
        pages = {1906-1921},
          doi = {10.1086/426006},
archivePrefix = {arXiv},
       eprint = {astro-ph/0406313},
 primaryClass = {astro-ph},
       adsurl = {https://ui.adsabs.harvard.edu/abs/2005AJ....129.1906C}
}

@ARTICLE{1986RMxAA..13..137A,
       author = {{Allen}, C. and {Martos}, M.~A.},
        title = "{A simple, realistic model of the galactic mass distribution for orbitcomputations.}",
      journal = {\rmxaa},
         year = 1986,
        month = dec,
       volume = {13},
        pages = {137-147},
       adsurl = {https://ui.adsabs.harvard.edu/abs/1986RMxAA..13..137A}
}









\bsp	
\label{lastpage}
\end{document}